%% file: main.tex
\documentclass[10pt,journal,compsoc]{IEEEtran}
\pdfoutput=1

\newcommand\preprint{}

\usepackage{amsmath,graphicx,amsfonts,amssymb}
\usepackage[ruled,boxed,vlined]{algorithm2e}
\usepackage[dvipsnames]{xcolor}
\usepackage[nocompress]{cite}
\usepackage{bm}
\usepackage{dsfont}
\usepackage{cases}

\usepackage{multirow}
\usepackage{array}
\newcolumntype{C}{>{\centering\arraybackslash}p{1.1cm}}

\usepackage[implicit=false,hidelinks]{hyperref}
\usepackage{comment}

\newcommand{\red}[1]{{\color{red} #1}}
\newcommand{\blue}[1]{{\color{blue} #1}}
\newcommand{\specialcell}[2][l]{%
  \begin{tabular}[#1]{@{}c@{}}#2\end{tabular}}
  
\newcommand{\matr}[1]{{\bm{#1}}}
\newcommand{\vect}[1]{{\bm{#1}}}
\newcommand{\norm}[1]{\left\lVert#1\right\rVert}
\newcommand{\argmin}{\mathop{\mathrm{arg\,min}}\limits}
\newcommand{\argmax}{\mathop{\mathrm{arg\,max}}\limits}
\newcommand{\prox}{\mathrm{prox}}

\newcommand{\netCst}{\textit{DRUNet-cst}}
\newcommand{\netVar}{\textit{DRUNet-var}}
\newcommand{\netVarCol}{\textit{DRUNet-var-RBG}}
\newcommand{\netDem}{\textit{DRUNet-dem}}

\newcommand{\hasappendix}[2]{\ifdefined\generateAppendix{#1}\else{#2}\fi}

\ifdefined\preprint
\newcommand\generateAppendix{}
\else
\newcommand\generateBiographies{}
\fi

\begin{document}

\title{Preconditioned Plug-and-Play ADMM\\ with Locally Adjustable Denoiser\\ for Image Restoration}
\author{Mikael~Le Pendu
        and~Christine~Guillemot

\thanks{
The authors are with
Inria Rennes -- Bretagne-Atlantique \\ 	  
263 Avenue G\'en\'eral Leclerc, 35042 Rennes Cedex, France (e-mail: firstname.lastname@inria.fr).}%
}

\IEEEtitleabstractindextext{%
\begin{abstract}
Plug-and-Play optimization recently emerged as a powerful technique for solving inverse problems by plugging a denoiser into a classical optimization algorithm. The denoiser accounts for the regularization and therefore implicitly determines the prior knowledge on the data, hence replacing typical handcrafted priors.
In this paper, we extend the concept of plug-and-play optimization to use denoisers that can be parameterized for non-constant noise variance. In that aim, we introduce a preconditioning of the ADMM algorithm, which mathematically justifies the use of such an adjustable denoiser. We additionally propose a procedure for training a convolutional neural network for high quality non-blind image denoising that also allows for pixel-wise control of the noise standard deviation. We show that our pixel-wise adjustable denoiser, along with a suitable preconditioning strategy, can further improve the plug-and-play ADMM approach for several applications, including image completion, interpolation, demosaicing and Poisson denoising.

\end{abstract}

}

\maketitle

\section{Introduction}

Inverse problems arising in image restoration require the use of prior knowledge on images in order to determine the most likely solutions among an infinity of possibilities. Traditional optimization methods rely on priors modeled as convex regularization functions such as the total variation, encouraging smoothness, or the $l_1$ norm for sparsity. In this context, convex optimization algorithms have played an important role and their convergence properties have been well-established.
However, the methods based on "hand-crafted" convex priors are now significantly outperformed by deep neural networks that directly learn an inverse mapping from the degraded measurements to the solution space.
Here, the prior knowledge on images and the degradation to recover from do not need a formal mathematical definition. Instead, they are implicitly taken into account when training the network from a large dataset of degraded images (i.e. network's intput) along with their original versions (i.e. network's output).
This approach has enabled a significant leap in performance for common image restoration problems including denoising, demosaicing, super-resolution, etc.
However, different neural networks must be carefully designed and trained for each problem specifically. Furthermore, it is usually impossible to interpret what task is performed by the different layers or sub-modules of the trained network which acts as a black box.

In order to increase the interpretability and genericity of deep neural networks, the field is evolving towards methods that combine them with traditional optimization algorithms. A popular approach, seen for example in \cite{diamond2017unrolled, Mardani2018, yang2016deep, gilton2019neumann, ning2020accurate,UnivDen18, DPDNN}, consists in defining so-called ``unrolled'' neural networks that simulate several iterations of an optimization algorithm.
For instance, in unrolled gradient descent, the gradient of the data-fidelity term can be computed knowing its closed form expression, while a convolutional neural network (CNN) is used to simulate the gradient computation of a regularization term. End-to-end training of this regularizing network is then performed by ``unrolling'' a given number of iterations of the optimization algorithm. Here, the trained CNN is meant to represent only prior knowledge on images, which does not depend on a specific image restoration task. However, in practice these networks require re-training for each task, meaning that task-specific features are also learnt in the end-to-end unrolled training.

A more universal approach referred to as ``Plug-and-Play'' (PnP), relies on proximal algorithms such as the Alternating Direction Method of Multipliers (ADMM) \cite{ADMM} or Half Quadratic Splitting (HQS) \cite{HQS}, where the proximal operator of the regularization term is usually replaced by a denoiser as in \cite{Venkatakrishnan13, zhang2021plug, zhang2017learning, OneNetwork, PoissonDen}. While Venkatakrishnan et al. \cite{Venkatakrishnan13} introduced this approach using traditional denoising techniques including BM3D \cite{BM3D} or K-SVD \cite{KSVD}, more recent methods attain higher performance thanks to neural networks trained for denoising. The main limitation, however, is that the objective function to minimize is typically non-convex when the proximal operator of the regularization term is defined as a high performance denoiser. Hence the proximal algorithms used may not provide the optimal solution.

In this paper, we propose a preconditioned ADMM in order to improve the algorithm's performance in this challenging ``Plug-and-Play'' scenario. While the proposed formulation remains mathematically equivalent to the original problem, the preconditioning makes it possible to use a denoiser that takes into account spatially varying noise levels. This is particularly advantageous for applications where the variance of the approximation error made at each iteration varies spatially in a predictable way. For instance, in image demosaicing, completion, or interpolation, only the unknown pixels that require interpolation are expected to have an approximation error. Hence, the pattern of known and unknown pixels can be used in our scheme to define a sensible preconditioning matrix. Likewise, when an image is corrupted by Poisson noise, the noise variance at each pixel is known to be proportional to the noiseless pixel intensity. In summary, the main contributions of our paper are as follows:
\begin{itemize}
    \item We formulate a preconditioned ADMM, which enables the use of a denoiser that can take into account spatially varying standard deviation.
    \item We propose a procedure for training a denoising neural network that takes as input a map of pixel-wise standard deviation, along with the image to be denoised.
    \item We demonstrate the effectiveness of the approach and define suitable preconditioning schemes for several applications: image completion, interpolation, demosaicing and Poisson denoising. Note that for Poisson denoising, we present further derivations accounting for the negative log-likelihood of the Poisson distribution, which replaces the conventional least square data fidelity term (only suitable for Gaussian noise).
\end{itemize}

\section{Background: Plug-and-Play ADMM}

\subsection{Plug-and-Play Prior}
\label{ssec:PnP}

Let us consider the linear inverse problem of recovering a clean image $\vect{\hat{x}}\in\mathbb{R}^n$ from its degraded measurements \mbox{$\vect{b}\in\mathbb{R}^m$} obtained with the degradation model $\vect{b} = \matr{A}\vect{x} + \vect{\nu}$, where \mbox{$\matr{A}\in\mathbb{R}^{m\times n}$} is the degradation matrix, $\vect{\nu}\in\mathbb{R}^m$ is additive white Gaussian noise with standard deviation $\sigma$, and $\vect{x}$ is the unknown ground truth image arranged as a vector. The problem is generally formulated as the Maximum a Posteriori (MAP) estimation of $\vect{x}$, given its prior probability density function $p$. The MAP is given by:
\begin{align}
\label{eq:MAP}
\vect{\hat{x}} &= \argmax_{\vect{x}} p(\vect{x}|\vect{b}) = \argmax_{\vect{x}} p(\vect{b}|\vect{x})\cdot p(\vect{x}),\\
&= \argmin_{\vect{x}} -\ln(p(\vect{b}|\vect{x})) -\ln(p(\vect{x})),\\
\label{eq:RLSQ}
&=\argmin_{\vect{x}} \frac{1}{2}\norm{\matr{A}\vect{x}-\vect{b}}_2^2 + \sigma^2\mathcal{R}(\vect{x}).
\end{align}
In practice, the prior distribution $p$ is not used directly, but is represented by the regularizer \mbox{$\mathcal{R}(\vect{x})=-\ln(p(\vect{x}))$}, which penalizes unlikely solutions.

The PnP approach introduced in \cite{Venkatakrishnan13} goes one step further by removing the need for an explicit definition of the regularizer. This is made possible by the use of proximal algorithms such as ADMM in which the regularizer only appears in the evaluation of the proximal operator of $\gamma\mathcal{R}$ for some scalar factor $\gamma$. This operator is defined as:
\begin{equation}
\prox_{\gamma\mathcal{R}}(\vect{u}) = \argmin_{\vect{x}}{\frac{1}{2}\norm{\vect{x}-\vect{u}}_2^2 + \gamma\cdot\mathcal{R}(\vect{x})}.
\end{equation}
It can be noted that the expression of $\prox_{\gamma\cdot\mathcal{R}}$ is a particular case of Eq.~\eqref{eq:RLSQ}, where the degradation matrix $\matr{A}$ is the identity matrix, i.e. the degradation model only consists in the addition of white Gaussian noise. In other words $\prox_{\gamma\cdot\mathcal{R}}$ is the MAP denoiser assuming additive white Gaussian noise of variance $\gamma$ and given the underlying prior $p(\vect{x})=e^{-\mathcal{R}(\vect{x})}$. Based on this observation, the authors of \cite{Venkatakrishnan13} replaced $\prox_{\gamma\cdot\mathcal{R}}$ by a standard image denoiser which implicitly determines the prior distribution.
 In the following, we will note the denoiser as a function of the image and the noise standard deviation:
\begin{equation}
\label{eq:gauss_denoise_fix}
F_\mathcal{R}(\vect{u},\sqrt\gamma) = \prox_{\gamma\mathcal{R}}(\vect{u})
\end{equation}

More recent works use a trained CNN to represent the proximal operator. For instance, in \cite{OneNetwork}, adversarial training is used to learn a projection operator. Here, the assumption is that the regularizer is the indicator function of the set of ``natural images''. Hence, the corresponding proximal operator projects the input image to the closest ``natural image'', which can be seen as a form of blind denoising. While this avoids the need for a parameter $\gamma$, it does not allow for controlling the denoising ``strength'' within the algorithm.
In \cite{zhang2021plug}, a neural network called DRUNet, that combines Res-Net \cite{Res-Net} and U-Net \cite{U-Net} architectures, is trained for the task of Gaussian denoising. The noise standard deviation is given as a constant input map in addition to the noisy image.
Hence, compared to approaches based on denoising CNNs trained for a single noise level\cite{Meinhardt17}, or for blind denoising \cite{OneNetwork, DnCNN, henz2018deep}, the method in \cite{zhang2021plug} can control the algorithm's parameters more precisely at each iteration, while keeping the highest denoising performance.
Since a single network is used, it also simplifies the method in \cite{zhang2017learning}, where 25 denoisers are trained for different standard deviations in order to cover a larger range of noise levels.

In this paper, we argue that a denoiser parameterized with an input map for standard deviation can even be trained to control the noise level at each pixel independently. Such a denoiser further improves the PnP scheme for several applications thanks to the proposed preconditioning.

\subsection{ADMM Formulation}
\label{ssec:ADMM}

In this section, we present the classical ADMM formulation in the case of the least squares inverse problem in Eq.~\eqref{eq:RLSQ}. The same notations will be used throughout the paper. In order to use the ADMM, the problem is first cast into a constrained optimization problem by splitting the two terms:
\begin{equation}\label{eq:prop1}
\begin{aligned}
\vect{\hat{x}} = &\argmin_{\vect{x},\vect{y}} & & \frac{1}{2} \norm{\matr{A}\vect{x}-\vect{b}}_2^2 + \sigma^2 \mathcal{R}(\vect{y})\\
&\text{subject to} & & \vect{x}=\vect{y},
\end{aligned}
\end{equation}

The constraint $\vect{x}=\vect{y}$ is included in the optimization by defining the augmented Lagrangian function $\mathcal{L}$, which introduces a dual variable $\vect{l}\in\mathbb{R}^n$ and a penalty parameter $\rho$:
\begin{align}
\begin{split}
    \mathcal{L}&(\vect{x},\vect{y},\vect{l}) = \frac{1}{2}\norm{\matr{A}\vect{x}-\vect{b}}_2^2 + \sigma^2\mathcal{R}(\vect{y}) \\
    &\qquad\qquad\qquad + \vect{l}^\mathsf{T} (\vect{x}-\vect{y}) + \frac{\rho}{2}\norm{\vect{x}-\vect{y}}_2^2,
\end{split}\\
    =& \frac{1}{2}\norm{\matr{A}\vect{x}-\vect{b}}_2^2 + \sigma^2\mathcal{R}(\vect{y}) + \frac{\rho}{2}\norm{\vect{x}-\vect{y}+\frac{\vect{l}}{\rho}}_2^2 - \frac{\norm{\vect{l}}_2^2}{2\rho}.
\end{align}

The ADMM method consists in alternatively minimizing the augmented Lagrangian $\mathcal{L}$ for the variables $\vect{x}$ and $\vect{y}$ separately, and by updating the dual variable $\vect{l}$. In practice, the penalty parameter $\rho$ may also be updated at each iteration to accelerate the convergence. The ADMM iteration thus reads as:
\begin{align}
\label{eq:data_subpb}
\vect{x^{k+1}} &= \argmin_{\vect{x}} \norm{\matr{A}\vect{x}-\vect{b}}_2^2 + \rho^k\norm{\vect{x}-\left(\vect{y^k}-\frac{\vect{l^k}}{\rho^k}\right)}_2^2,\\
\label{eq:prior_subpb}
\vect{y^{k+1}} &= \argmin_{\vect{y}} \frac{1}{2}\norm{\vect{y}-\left(\vect{x^{k+1}}+\frac{\vect{l^k}}{\rho^k}\right)}_2^2 + \frac{\sigma^2}{\rho^k}\mathcal{R}(\vect{y}),\\
\vect{l^{k+1}} &= \vect{l^k} + \rho^k(\vect{x^{k+1}}-\vect{y^{k+1}}),\\
\label{eq:rho-decrease}
\rho^{k+1} &= \rho^k \cdot \alpha, \quad\text{with}\quad \alpha > 1,
\end{align}
where the dual variable $\vect{l}$ is typically zero-initialized. An initialization of $\vect{x}$ is also required and is often performed as $\vect{x}_0=\matr{A}^\mathsf{T}\vect{b}$, but a finer initialization strategy may be used depending on the problem.

The y-update in Eq.~\eqref{eq:prior_subpb} can be equivalently written using a proximal operator:
\begin{equation}
    \vect{y^{k+1}} = \prox_{\gamma\mathcal{R}}(\vect{x^{k+1}}+\vect{l^k}/\rho^k),
\end{equation}
where $\gamma=\sigma^2/\rho^k$. As seen in Section \ref{ssec:PnP}, this step can be performed in the PnP approach by denoising the image \mbox{$\vect{x^{k+1}}+\vect{l^k}/\rho^k$} using a Gaussian denoiser for a noise standard deviation $\sqrt{\gamma} = \sigma/\sqrt{\rho^k}$.

\section{Preconditioned Plug-and-Play ADMM}
\label{sec:PrecoPnpADMM}

In several problems, we have additional knowledge about locally varying noise level. For instance, in image completion or interpolation problems, the pixels already sampled in $\vect{b}$ are known without error. Thus, at a given iteration, only the pixels at unknown positions need to be denoised. However, in the original ADMM formulation the denoising step considers the same noise level $\sigma/\sqrt{\rho^k}$ at each pixel.
Hence, in order to finely tune the denoising effect and improve the performance of the PnP-ADMM, we reformulate the problem with a preconditioner that takes into account the knowledge of a variable noise level.

\subsection{ADMM Reformulation}
Let us consider the following problem, mathematically equivalent to Eq.~\eqref{eq:RLSQ}:
\begin{eqnarray}
\label{eq:PRLSQ}
    \vect{\Tilde{x}} &=& \argmin_{\vect{x}} \frac{1}{2}\norm{\matr{AP}\vect{x}-\vect{b}}_2^2 + \sigma^2\mathcal{R}(\matr{P}\vect{x}),\\
\label{eq:PRLSQ2}
    \vect{\hat{x}} &=& \matr{P}\vect{\Tilde{x}},
\end{eqnarray}
where $\matr{P}$ is a diagonal preconditioning matrix.

Similarly to the Section.~\ref{ssec:ADMM}, we can express the ADMM algorithm for solving Eq.~\eqref{eq:PRLSQ} by replacing the updates of $\vect{x}$ and $\vect{y}$ in Eqs.~\eqref{eq:data_subpb} and~\eqref{eq:prior_subpb} with respectively:
\begin{align}
\label{eq:data_subpb_preco}
    \vect{x^{k+1}} &= \argmin_{\vect{x}} \norm{\matr{AP}\vect{x}-\vect{b}}_2^2 + \rho^k\norm{\vect{x}-\left(\vect{y^k}-\frac{\vect{l^k}}{\rho^k}\right)}_2^2,\\
\label{eq:prior_subpb_preco}
     \vect{y^{k+1}} &= \argmin_{\vect{y}} \frac{1}{2}\norm{\vect{y}-\left(\vect{x^{k+1}}+\frac{\vect{l^k}}{\rho^k}\right)}_2^2 + \frac{\sigma^2}{\rho^k}\mathcal{R}(\matr{P}\vect{y}).
\end{align}

The variable $\vect{x}$ can be updated directly using the well-known closed form solution of Eq.~\eqref{eq:data_subpb_preco}:
\begin{equation}
\label{eq:data_subpb_solution_preco}
    \vect{x^{k+1}} = (\matr{P^{\top}A^{\top}AP}+\rho^k \matr{I})^{-1}(\matr{P^{\top}A^{\top}} \vect{b} + \rho^k \vect{y^k} - \vect{l^k}).
\end{equation}
However, for updating $\vect{y}$, the denoiser $F_\mathcal{R}$ from Eq.~\eqref{eq:gauss_denoise_fix}, is no longer suitable because of the matrix $\matr{P}$ within the regularization term. The next section presents how the y-update can still be performed using a more general denoiser that considers Gaussian noise with variable variance.

\subsection{New Prior Term Sub-Problem}
\label{ssec:reg_preco_prox}

\subsubsection{General Gaussian Denoiser}
Let us first define the general expression of a denoiser for Gaussian noise with covariance matrix $\matr{\Sigma}$. The MAP estimate from a noisy image $\vect{u}$ can be derived similarly to Eqs.~\eqref{eq:MAP}-\eqref{eq:RLSQ}, where the likelihood $p(\vect{u}|\vect{x})$ is a multivariate Gaussian distribution, i.e. \mbox{$p(\vect{u}|\vect{x})\propto e^{-\frac{1}{2}\norm{\matr{\Sigma}^{-1/2}(\vect{u}-\vect{x})}_2^2}$}. Including the prior and taking the negative logarithm gives the expression of the denoiser:
\begin{equation}
    \label{eq:gauss_denoise_var}
    G_\mathcal{R}(\vect{u},\matr{\Sigma}^{1/2}) = \argmin_\vect{x} \frac{1}{2}\norm{\matr{\Sigma}^{-1/2}(\vect{u}-\vect{x})}_2^2 + \mathcal{R}(\vect{x}),
\end{equation}
Here, we will only consider the case of a diagonal matrix $\matr{\Sigma}$. Hence, we still assume independent noise at each pixel (but not identically distributed), and the diagonal of $\matr{\Sigma}$ corresponds to pixel-wise variance. Note that the denoiser $F_\mathcal{R}$ is equivalently defined as \mbox{$F_\mathcal{R}(\vect{x},\sigma) = G_\mathcal{R}(\vect{x},\sigma \matr{I})$} for a constant standard deviation $\sigma$. A practical image denoiser $G_\mathcal{R}$ can be obtained by training a CNN that takes as input a map of the pixel-wise noise levels in addition to the noisy image. A training procedure for such a denoiser is presented in Section~\ref{sec:DeepDenoiser}.

\subsubsection{Sub-Problem Solution}
Now, let us rewrite Eq.~\eqref{eq:prior_subpb_preco} using the change of variable \mbox{$\vect{\Tilde{y}} = \matr{P}\vect{y}$}, which gives \mbox{$\vect{y} = \matr{P}^{-1}\vect{\Tilde{y}}$}. Thus, we have \mbox{$\vect{y^{k+1}} = \matr{P}^{-1}\vect{\Tilde{y}^{k+1}}$} and \mbox{$\vect{\Tilde{y}^{k+1}} = \argmin_{\vect{\Tilde{y}}} f(\vect{\Tilde{y}})$} with:
\begin{align}
    f(\vect{\Tilde{y}})&= \frac{\rho^k}{2\sigma^2}\norm{\matr{P}^{-1}\vect{\Tilde{y}}-\left(\vect{x^{k+1}}+\frac{\vect{l^k}}{\rho^k}\right)}_2^2 + \mathcal{R}(\vect{\Tilde{y}}),\\
    &=\frac{1}{2}\norm{\frac{\sqrt{\rho^k}}{\sigma}\matr{P}^{-1}\left(\vect{\Tilde{y}}-\matr{P}\left(\vect{x^{k+1}}+\frac{\vect{l^k}}{\rho^k}\right)\right)}_2^2+\mathcal{R}(\vect{\Tilde{y}}).
\end{align}
Given the denoiser $G_\mathcal{R}$ in Eq.~\eqref{eq:gauss_denoise_var} we can then rewrite the y-update step as:
\begin{equation}
\label{eq:y-update-preco}
    \vect{y^{k+1}} = \matr{P}^{-1}G_\mathcal{R}\left(\matr{P}\left(\vect{x^{k+1}}+\frac{\vect{l^k}}{\rho^k}\right), \; \frac{\sigma}{\sqrt{\rho^k}}\matr{P}\right)
\end{equation}

Therefore, the preconditioned ADMM still solves the original problem but introduces a denoising step that assumes  noise with spatially varying standard deviation. The standard deviation can be adjusted per pixel via the diagonal matrix $\matr{P}$ and the global parameters $\sigma$ and $\rho^k$. A suitable choice of the preconditioning matrix can be made depending on the problem in order to improve the performance of the PnP-ADMM.

\subsection{Variables Interpretation and Initialization}
\label{ssec:init_preco}
It should be noted that in the preconditioned problem, an image $\vect{x^k}$ only estimates the intermediate variable $\vect{\tilde{x}}$, but the final result is $\vect{\hat{x}}=\matr{P}\vect{\tilde{x}}$ (see Eq.~\eqref{eq:PRLSQ2}). Hence, $\matr{x^k}$ (similarly $\vect{y^k}$) may not be a plausible image. However, the input of the denoiser is $\matr{P}(\vect{x^{k-1}}+\vect{l^{k}}/\rho^k)$, where the multiplication by $\matr{P}$ rescales the image pixels consistently with the ``natural image'' $\vect{\hat{x}}$, as expected by the denoiser.

Inversely, for the initialization, the inverse scaling must be performed. Given an initial estimate $\vect{\hat{x}^0}$ of the true image $\vect{\hat{x}}$, a good initialization for our problem is thus \mbox{$\vect{x^0}=\matr{P}^{-1}\vect{\hat{x}^0}$}.

\section{Deep Locally Adjustable Denoiser}
\label{sec:DeepDenoiser}
\subsection{Denoising Network for Known Noise Level}

Our algorithm relies on a deep neural network for solving the prior term sub-problem which corresponds to a Gaussian denoising problem where the noise level (i.e. standard deviation) is known and can be adjusted for each pixel.
Here, we use the DRUNet architecture proposed in~\cite{zhang2021plug}. The overall structure consists of a U-Net where each level contains sequences of 4 residual-blocks and either a down-sampling layer (first branch of the U) or an upsampling layer (second branch).

The DRUNet network structure in \cite{zhang2021plug} conforms to the definition of the denoisers $F_\mathcal{R}$ (Eq.~\eqref{eq:gauss_denoise_fix}) or $G_\mathcal{R}$ (Eq.~\eqref{eq:gauss_denoise_var}) by taking as input the concatenation (in the channel dimension) of the noisy image and a noise level map whose pixels' values are equal to the noise standard deviation. The advantage of using an input noise level map is that a single generic model can be trained to perform optimally for a large range of noise levels. 
However, in \cite{zhang2021plug} the model is only trained considering constant noise level maps since their algorithm only uses the constant denoiser $F_\mathcal{R}$. In the rest of the paper, we refer to this network as \netCst{}.

In order to use our preconditioning approach, a locally adjustable denoiser $G_\mathcal{R}$ is required. Therefore, the training process must be adapted so that each input sample consists of an arbitrary noise level map along with an image corrupted with the corresponding Gaussian noise level for each pixel.
We describe in the next section how to randomly generate suitable patterns so that the trained model generalizes well for any arbitrary noise level map.

\subsection{Noise Level Map Generation}
\label{ssec:noiselevelmap}

Let us first consider the case of a constant noise level map as in \cite{zhang2021plug}. Here, all the pixels are equal to the same random variable $S$ that can be simply defined as:
\begin{equation}
\label{eq:cstSigma}
    S = 2\mu\cdot X,
\end{equation}
where $X$ is a uniformly distributed random variable in the range $[0,1]$.
The parameter $\mu$ is equal to the expectation of the random variable $S$. It can be selected to train a denoiser that is sufficiently generic for all the noise levels $\sigma$ in the range [0,$2\mu$].

Now, in order to train a locally adjustable denoiser $G_\mathcal{R}$, we generate a random map for each training image using the following random process:
\begin{equation}
\label{eq:randSigma}
    S_i = 2\mu\cdot(X_i\cdot(1-W) + O\cdot W),
\end{equation}
where $i$ is the pixel index, $O$, $W$ and $X_i$ (for each pixel $i$) are independent random variables with uniform distributions in the range $[0,1]$. One can verify that $\mu$ is the expectation of $S_i$, and the range of possible values of $S_i$ is [0,$2\mu$] similarly to the previous case with constant noise level.

The random weight $W$ allows us to adjust the variance of the noise level map (i.e. lower weight corresponding to higher variance).
The random offset $O$ is also necessary to reduce the correlation between the variance and the mean of the generated maps. In particular, the offset enables the generation of maps with a low variance but a high average level, so that the trained denoiser generalizes well even in the constant noise level case.
Hence, the method makes it possible to train the network with noise level maps covering a wide range of first and second order statistics, which prevents overfitting for a specific type of pattern. In the paper, we refer to the network trained with this method as \netVar{}.

For some applications, the noise standard deviation may also need to be adjusted depending on the color component. For these applications, we train another network called \netVarCol{} that takes an input noise level map for each color component. For the training, the noise level maps are generated for each component separately using the random process in Eq.~\eqref{eq:randSigma}.

\begin{figure*}[t]
\centering
\begin{minipage}[h]{.194\linewidth}
	\centerline{\includegraphics[width=\linewidth]{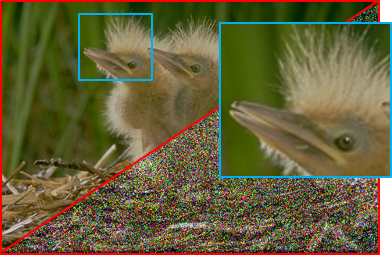}}
	\centerline{\small{(a) Ground truth / noisy}}
	\centerline{\small{($\mu=100/255$)}}
\end{minipage}
\begin{minipage}[h]{.194\linewidth}
	\centerline{\includegraphics[width=\linewidth]{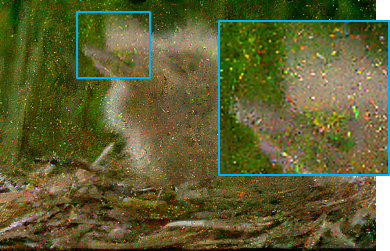}}
	\centerline{\small{(b) \textit{Avg-std}}}
	\centerline{\small{PSNR=19.18}}
\end{minipage}
\begin{minipage}[h]{.194\linewidth}
	\centerline{\includegraphics[width=\linewidth]{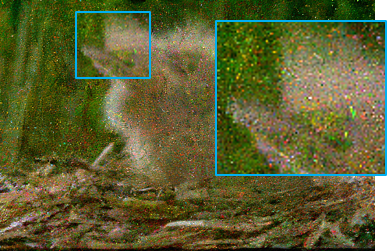}}
	\centerline{\small{(c) \textit{True-std}}}
	\centerline{\small{PSNR=19.07}}
\end{minipage}
\begin{minipage}[h]{.194\linewidth}
	\centerline{\includegraphics[width=\linewidth]{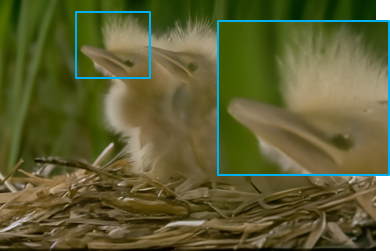}}
	\centerline{\small{(d) \textit{PnP-ADMM}}}
	\centerline{\small{PSNR=28.94}}
\end{minipage}
\begin{minipage}[h]{.194\linewidth}
	\centerline{\includegraphics[width=\linewidth]{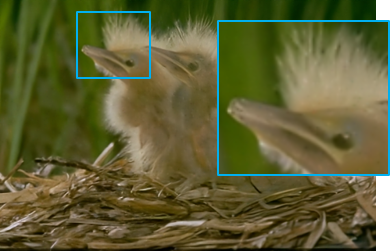}}
	\centerline{\small{(e) \netVar{}}}
	\centerline{\small{PSNR=29.28}}
\end{minipage}
\vspace{-1pt}
\caption{Denoising results with pixel-wise standard deviation randomly selected in $[0,2\mu]$. The average noise standard deviation is $\mu=100/255$. For the \textit{Avg-std}, \textit{True-std} and \textit{PnP-ADMM} schemes in (b-d), the network \netCst{} is used. In (e), the network \netVar{} is used with the \textit{True-std} scheme.}
\label{fig:den-var}
\vspace{-10pt}
\end{figure*}

\subsection{Training Details}

The training is performed by generating a noise level map randomly for each input image, as described in Section~\ref{ssec:noiselevelmap}. Considering pixel data in the range $[0,1]$, we use the parameter $\mu=25/255$ in Eq.~\eqref{eq:cstSigma} (for the \netCst{} denoiser) or Eq.~\eqref{eq:randSigma} (for the \netVar{} and \netVarCol{} denoisers). The maximum noise level is thus $2\mu=50/255$. Gaussian noise with standard deviation defined by the noise level map is then added to the input image.

The remaining details of the training procedure are the same as described in \cite{zhang2021plug}. We use the same large dataset of 8694 images, including the Berkeley Segmentation Dataset BSDS500 (without the validation images)~\cite{BSDS500}, the Waterloo Exploration Database~\cite{WaterlooDataset}, the DIV2K dataset~\cite{DIV2K}, and the Flickr2K dataset~\cite{Flickr2K}. We minimize the $l_1$ loss between the denoiser's output and the ground truth image using the ADAM optimizer~\cite{ADAM} with a learning rate initialized at $1e-4$ and decreased by half every 100000 iterations. The training is stopped when the learning rate is smaller than $5e-7$. Finally, each iteration of the training uses a batch of 16 patches of size 128x128 randomly selected from the images of the training dataset.

\subsection{Denoising Performance}

Here, we evaluate the performance of our denoisers \netVar{} and \netVarCol{} generalized for variable standard deviation, in comparison with \netCst{}~\cite{zhang2021plug} as well as other state-of-the-art denoisers that assume a constant noise level. These include the BM3D denoiser (noted CBM3D for color images)~\cite{BM3D} and the two recent CNN-based methods FFDNet~\cite{FFDNET} and RNAN~\cite{RNAN}. Similarly to \netCst{}, the FFDNet denoiser can be parameterized at inference time knowing the noise standard deviation. On the other hand, RNAN requires a separate training for each noise level, but has shown higher denoising performance thanks to a non-local attention module.

We perform our evaluations with the widely used Set5~\cite{Set5} and CBSD68~\cite{CBSD68} test datasets.

\subsubsection{Constant Noise Level}
First, let us consider input images corrupted with constant noise level. The results in Table~\ref{tab:den-cst} show that our locally adjustable denoisers \netVar{} and \netVarCol{} do not have significantly degraded performance compared to the reference network \netCst{}, despite being trained for more generic noise level maps. Only a very moderate loss is observed for the \netVarCol{} denoiser compared to \netCst{}, mostly for the lowest and highest noise levels ($\sigma=10$ or $50$). Nevertheless, \netVarCol{} still outperforms the other state-of-the-art methods, even the RNAN denoiser trained specifically for each noise level.

\begin{table}
\setlength\tabcolsep{5.5pt}
\centering
\caption{Denoising performance (average PSNR over each dataset) for Gaussian noise with constant noise level (standard deviation $\sigma=10$, $30$ or $50$).}
\vspace{-10pt}
\begin{tabular}{r|ccc|ccc|}
\cline{2-7}
Dataset                  & \multicolumn{3}{c|}{CBSD68} & \multicolumn{3}{c|}{Set5} \\ \cline{2-7} 
\multicolumn{1}{r|}{noise level $\sigma$}
                                            &  10   &  30   &  50   &  10   &  30   &  50   \\ \hline
\multicolumn{1}{|l|}{CBM3D~\cite{BM3D}}     & 35.90 & 29.91 & 27.51 & 36.02 & 30.97 & 28.75 \\
\multicolumn{1}{|l|}{FFDNet~\cite{FFDNET}}  & 36.14 & 30.32 & 27.97 & 36.16 & 31.35 & 29.24 \\
\multicolumn{1}{|l|}{RNAN~\cite{RNAN}}      & 36.43 & 30.65 & 28.30 & 36.62 & 31.77 & 29.61 \\ \hline
\multicolumn{1}{|l|}{\netCst{}~\cite{zhang2021plug}}
                                            & 36.51 & 30.79 & 28.48 & 36.67 & 31.90 & 29.80 \\
\multicolumn{1}{|l|}{\netVar{}}             & 36.51 & 30.80 & 28.48 & 36.67 & 31.91 & 29.79 \\
\multicolumn{1}{|l|}{\netVarCol{}}          & 36.47 & 30.78 & 28.46 & 36.61 & 31.90 & 29.77 \\ \hline
\end{tabular}
\label{tab:den-cst}
\vspace{-10pt}
\end{table}

\subsubsection{Spatially Varying Noise Level}

\begin{table}
\centering
\renewcommand*{\arraystretch}{1.5}
\setlength\tabcolsep{5pt}
\caption{Denoising performance (average PSNR over each dataset) for Gaussian noise with pixel-wise variable standard deviation in the range $[0,2\mu]$.}
\vspace{-10pt}
\begin{tabular}{cc ccc|c}
& & \multicolumn{3}{c}{\netCst{}~\cite{zhang2021plug}} &    \netVar{}   \\ \cline{3-6}  & \multicolumn{1}{l|}{Dataset}  & \textit{Avg-std} & \textit{True-std} & \textit{PnP-ADMM} & \multicolumn{1}{c|}{\textit{True-std}} \\ \cline{2-6}
\multicolumn{1}{c|}{\multirow{2}{*}{\parbox{9mm}{\rotatebox{45}{$\mu=\frac{25}{255}$}}}} & \multicolumn{1}{c|}{Set5}  & 29.01 & 29.03 & 34.35 &\multicolumn{1}{c|}{ 34.48 }\\
\multicolumn{1}{c|}{} & 
\multicolumn{1}{c|}{BSD68} & 28.14 & 28.22 & 32.75 &\multicolumn{1}{c|}{ 32.90 }\\\cline{2-6}
\multicolumn{1}{c|}{\multirow{2}{*}{\parbox{9mm}{\rotatebox{45}{$\mu=\frac{50}{255}$}}}} & \multicolumn{1}{c|}{Set5}  & 24.45 & 24.37 & 32.03 &\multicolumn{1}{c|}{ 32.24 }\\
\multicolumn{1}{c|}{} & 
\multicolumn{1}{c|}{BSD68} & 23.59 & 23.60 & 30.09 &\multicolumn{1}{c|}{ 30.27 }\\ \cline{2-6}
\multicolumn{1}{c|}{\multirow{2}{*}{\parbox{9mm}{\rotatebox{45}{$\mu=\frac{100}{255}$}}}} & \multicolumn{1}{c|}{Set5}  & 19.82 & 19.58 & 29.69 &\multicolumn{1}{c|}{ 30.06 }\\
\multicolumn{1}{c|}{} & 
\multicolumn{1}{c|}{BSD68} & 19.03 & 18.88 & 27.73 &\multicolumn{1}{c|}{ 28.01 }\\ \cline{2-6}
\end{tabular}
\label{tab:den-variable-sigma}
\vspace{-10pt}
\end{table}

Now, let us show the advantage of the proposed training procedure by comparing the performances of our \netVar{} denoiser with the \netCst{} version when the input noise standard deviation varies spatially.
For these tests, we randomly select a standard deviation in the range $[0,2\mu]$ for each pixel. For a complete comparison with \netCst{}, we have tested this network in three different ways: 
\begin{itemize}
    \item \textit{Avg-std}: using the average noise level $\mu$ as input.
    \item \textit{True-std}: using the true noise level map as input.
    \item \textit{PnP-ADMM}: solving the problem of denoising with variable noise level (Eq.\eqref{eq:gauss_denoise_var}) using PnP-ADMM without preconditioning based on the \netCst{} network. 
\end{itemize}
For the \textit{PnP-ADMM}, we set the parameters using Eq.~\eqref{eq:params-setting} so that the noise of highest standard deviation \mbox{$\sigma^0_{den}=2\mu$} is removed at the first iteration, while successive iterations refine the result by decreasing the noise level of the denoiser down to $\sigma^N_{den}$. Here, we use $\sigma^N_{den}=\frac{2\mu}{3}$ for $N=25$ iterations, which we found to give the best results for this application.

Table~\ref{tab:den-variable-sigma} and Fig.~\ref{fig:den-var} present the results of \netCst{} in each of the three schemes, and \netVar{} in the \textit{True-std} scheme.
As expected, the results using only the average standard deviation are unsatisfying, since the pixels with above average noise level are not sufficiently denoised, while details are not optimally preserved in less noisy regions. The same behavior is observed when using the true noise level map as input to \netCst{}. This confirms that a denoiser trained for constant noise level does not generalize when the noise standard deviation strongly varies between pixels.
More satisfying results are obtained when using \netCst{} in the PnP-ADMM scheme. However, our \netVar{} denoiser trained directly for this task can still retrieve more details, which indicates that the PnP-ADMM without preconditioning is sub-optimal.
More realistic applications are presented in the rest of the paper to demonstrate that our locally adjustable denoisers also allow for improved performances in practical scenarios thanks to the preconditioned PnP-ADMM.

\section{Applications}
\label{sec:Applications}

In this section we present several examples of applications in image restoration. Although the proposed denoiser allows us to adjust of the variance locally, it still assumes an independent noise distribution for each pixel. Hence, in order to demonstrate the advantage of our preconditioning for the PnP-ADMM scheme, we restrict our study to applications where the error variance is expected to vary only in the pixel domain. While this excludes deblurring, or super-resolution from an anti-aliased downsampled image, many applications of high practical interest are still concerned. These include image completion, interpolation, demosaicing and Poisson denoising.

\vspace{-5pt}
\subsection{Image Completion and Interpolation}
\label{ssec:CompleteAndInterp}
\vspace{-2pt}
In the problems of image completion and interpolation, a subset $\Omega$ of the original image pixels are sampled in the input vector $\vect{b}$, while the remaining ones are unknown and must be determined. In both problems, the sampling can be expressed in the matrix form $\matr{A}\vect{x}=\vect{b}$ with a sampling matrix $\matr{A}$ such that each row contains only zeros, except one element equal to $1$ at the index of a sampled pixel.

When solving the problem with ADMM, all the unknown pixels are expected to have the highest estimation error at the initialization stage. However, over iterations, more reliable estimates should be obtained especially for pixels close to a sampled pixel. Hence, the denoiser within our preconditioned PnP-ADMM should be adjusted locally depending on the proximity to a sampled pixel. Following this intuition, we define a preconditioning matrix $\matr{P}$ that varies over iterations according to the following formula for each iteration $k$:
\begin{equation}
\label{eq:P_interp1}
    \left[\matr{P^k}\right]_{i,i} = \frac{\max(\vect{m^k})+\epsilon}{\left[\vect{m^k}\right]_i+\epsilon}
\end{equation}
where $\epsilon$ is a scalar parameter and $\vect{m^k}$ is defined recursively by blurring $\vect{m^{k-1}}$, and where $\vect{m^0}$ is the mask indicating the known pixels:
\begin{equation}
\label{eq:P_interp2}
    \vect{m^k} = \vect{m^{k-1}} * g(\sigma_f),
    \,\,\,\text{with} \,\,\,
    \left[\vect{m^{0}}\right]_i =
    \begin{cases}
    1 \quad\text{if}\quad i\in\Omega\\
    0 \quad\text{otherwise}
    \end{cases}
    \forall i
\end{equation}
Here, * is the 2D convolution operation and $g$ is a 2D Gaussian filter of parameter $\sigma_f$. Note that it is equivalent to have \mbox{$\vect{m^k}=\vect{m^0}*g(\sigma_f\sqrt{k})$}. We can thus determine $\sigma_f$ so that the preconditioning at the last iteration $N$ does not depend on $N$, but only on a fixed parameter $\sigma_f^{last}$ at the last iteration by taking $\sigma_f=\sigma_f^{last}/\sqrt{N}$.

Using this definition, the preconditioning values are always equal to $1$ for the known pixels (i.e. $\left[\matr{P^k}\right]_{i,i}=1\,\, \forall i\in\Omega$), and higher than $1$ for the other pixels, which results in a stronger denoising for the unknown pixels in Eq.~\eqref{eq:y-update-preco}. The parameter $\epsilon$ prevents too high preconditioning values that would make the denoising impractical. The highest preconditioning value is $p_{max}=(1+\epsilon)/\epsilon$.
Additionally, thanks to the Gaussian blur, the preconditioning values of unknown pixels will depend on their distance with sampled pixels. Therefore, the least reliable pixels (i.e. that are far from sampled pixels) will be denoised more.

Note that we only use non-zero preconditioning values (and thus non-zero standard deviation in the denoiser) so that the matrix $\matr{P}$ remains invertible. However, ideally, no denoising should be performed at the sampled pixel positions to prevent unnecessary loss of information. Therefore, we force the denoiser's output to be equal to the input for these pixels.

\vspace{-5pt}
\subsection{Demosaicing}
\label{ssec:demosaicing}
\vspace{-2pt}
Digital cameras typically capture colors using a sensor with a color filter array (CFA). Thanks to the mosaic formed by the CFA, neighbor pixels on the sensor record a different color information. Knowing the CFA pattern, the full color information can then be retrieved by an inverse problem called demosaicing.

Red, green and blue filters are generally used so that each pixel directly records one of the RGB components. Demosaicing can then be seen as an interpolation problem where a pixel R, G or B value is either known or unknown. Therefore, we use the same preconditioning strategy described in Section~\ref{ssec:CompleteAndInterp}. However, in a realistic scenario, sensor data may also contain noise which is preferably removed jointly with the demosaicing step \cite{gharbi2016deep}. In the case of noisy data, our method applies similarly, but we do not force the denoiser to keep the sampled pixels unchanged as explained in Section~\ref{ssec:CompleteAndInterp}.

Furthermore, unlike the problems of completion and interpolation, the demosaicing uses different masks indicating the sampled pixels for the R, G and B components. The preconditioning matrix thus takes different values for each component of the same pixel, which requires using a denoiser parameterized with a RGB noise level map. Our \netVarCol{} network trained for arbitrary noise level patterns is thus suitable for this application.

However, in order to illustrate the tradeoff between the genericity of the denoiser and the optimal performance, we also trained a more specialized denoiser for demosaicing that we call \netDem{}. For training this network, instead of generating noise level maps with the generic random process in Eq.~\eqref{eq:randSigma}, 
we generate the patterns from the CFA mask, that is also used in the preconditioned ADMM.
These patterns are defined by Eqs.~\eqref{eq:P_interp1} and~\eqref{eq:P_interp2}, where a mask $\vect{m^0}$ is defined for each color component by the CFA pattern.
Experimental results using either the generic network \netVarCol{} or the specialized one \netDem{}, are given in Section~\ref{ssec:resDem}.

\vspace{-5pt}
\subsection{Poisson Denoising}
\label{ssec:Poisson}
\vspace{-2pt}

In many practical scenarios, the assumption that images are corrupted by additive white Gaussian noise is inaccurate. Camera noise is typically better modelled by a Poisson process.
In order to formulate a MAP optimization for the problem of denoising an image corrupted with Poisson noise, the least squares data term in Eq.~\eqref{eq:RLSQ} must be modified. Here, we re-derive our preconditioned ADMM algorithm for Poisson denoising.

\subsubsection{Preconditioned ADMM for Poisson denoising}

Applying Poisson noise to an original image $\vect{x\in\mathbb{R}^n}$ gives a noisy image $\vect{b}\in\mathbb{N}^n$ such that the likelihood distribution is:
\begin{equation}
    p(\vect{b}|\vect{x}) = \prod_{i=1}^{n} \frac{\vect{x}_i^{\vect{b}_i}e^{-\vect{x}_i}}{\vect{b}_i!}.
\end{equation}

As seen in Eqs. \eqref{eq:MAP}-\eqref{eq:RLSQ}, the data term is the negative log-likelihood which can be derived as
\begin{equation}
    -\ln(p(\vect{b}|\vect{x})) = -\vect{b}^\mathsf{T}\ln(\vect{x}) + \mathds{1}^\mathsf{T}\vect{x} + c,
\end{equation}
where $c$ is a constant term that can be ignored for the minimization. The preconditioned Poisson denoising problem is then formulated by substituting $\vect{x}$ with $\matr{P}\vect{x}$ and by adding the regularization term as in Eq.~\eqref{eq:PRLSQ}:
\begin{equation}
\label{eq:PoissonDenPb}
    \vect{\Tilde{x}} = \argmin_{\vect{x}} -\vect{b}^\mathsf{T}\ln(\vect{\matr{P}x}) + \mathds{1}^\mathsf{T}\vect{\matr{P}x} + \mathcal{R}(\matr{P}\vect{x}),
\end{equation}

The problem is solved using the ADMM, by splitting the data and regularization terms following the methodology in Section~\ref{ssec:ADMM}. Since the regularization term remains the same as in Section~\ref{sec:PrecoPnpADMM}, the y-update in Eq.~\eqref{eq:y-update-preco} still applies, and is performed using the same Gaussian denoiser $G_\mathcal{R}$. However, given the new data term, the x-update becomes:
\begin{equation}
\label{eq:x-update-Poisson1}
\vect{x^{k+1}} = \argmin_\vect{x} -\vect{b}^\mathsf{T}\ln(\matr{P}\vect{x}) + \mathds{1}^\mathsf{T}\matr{P}\vect{x} + \frac{\rho^k}{2}\norm{\vect{x}-\vect{u^k}}_2^2,
\end{equation}
where $\vect{u^k}=\vect{y^k}-\vect{l^k}/\rho^k$.

Knowing that $\matr{P}$ is a diagonal matrix, this problem can be solved independently for each pixel. The following closed form solution is derived in \hasappendix{Appendix \ref{app:Poisson}}{the supplementary materials} (similar derivations are also found in \cite{PoissonDen} for the case $\matr{P}=\matr{I}$):
\begin{equation}
\label{eq:x-update-Poisson2}
\left[\vect{x^{k+1}}\right]_i = \frac{\rho^k\vect{u}_i-\matr{P}_{i,i}+\sqrt{(\rho^k\vect{u}_i-\matr{P}_{i,i})^2+4\rho^k\vect{b}_i}}{2\rho^k}.
\end{equation}

It is worth noting that unlike the Gaussian case, the noise level only depends on the data and is not controlled by a parameter $\sigma$. However, in photography the amount of Poisson noise can be reduced by increasing the exposure. This effect can be simulated by applying Poisson noise to a ground truth image rescaled in the range $[0,\lambda]$. A high peak value $\lambda$ thus simulates a high exposure and less noise.
However, for an input image $\vect{b}$ with values in the range $[0,\lambda]$, the algorithm must be modified since the denoiser $G_\mathcal{R}$ in Eq.~\eqref{eq:y-update-preco} assumes data in the range $[0,1]$.
The y-update should then be performed by replacing $G_\mathcal{R}$ with the denoiser $G^\lambda_\mathcal{R}$ adapted to the range $[0,\lambda]$ by rescaling the input and output as:
\begin{equation}
    G^\lambda_\mathcal{R}(\vect{u},\matr{\Sigma}^{1/2}) = \lambda G_\mathcal{R}(\vect{u}/\lambda,\matr{\Sigma}^{1/2}/\lambda)
\end{equation}
Similarly, the final ADMM result should be divided by $\lambda$ to recover an image in the range $[0,1]$.

\subsubsection{Choice of Preconditioning Matrix}

It is well known that the variance and the expected value of a random variable with the Poisson distribution are both equal to the distribution's parameter, i.e. the noiseless image pixel. This means that for each pixel of the image, the noise standard deviation is equal to the square root of the pixel value in the ground truth image. Although this image is unknown, the noisy image provides a first estimate of the noise variance, which can be refined at the next iterations, as we obtain a better estimate of the noiseless image.

Therefore, in this problem, we use $\rho_k=1$ for each iteration, so that the denoiser is only controlled using the preconditioning matrix which is initialized as \mbox{$\matr{P^0}=\mathrm{diag}(\sqrt\vect{b})$}. Here, the square root is applied element-wise and the notation $\mathrm{diag}$ forms a diagonal matrix from the vector.
Following the initialization strategy in Section~\ref{ssec:init_preco}, the variable $\vect{x}$ is then initialized such that \mbox{$\vect{\hat{x}^0}=\matr{P^0}\vect{x^0}=\vect{b}$}, and thus \mbox{$\vect{x^0}=\left[\matr{P^0}\right]^{-1}\vect{b}=\sqrt{\vect{b}}$}.

Then, at each iteration, a more accurate estimate of the noiseless image is given by $\matr{P^k}\vect{y^k}$ obtained after the denoising step in Eq.~\eqref{eq:y-update-preco}. Note that $\matr{P^k}\vect{y^k}$ is directly the output of the denoiser (before applying $\left[\matr{P^k}\right]^{-1}$)
and is thus obtained without further matrix multiplication. Hence, the preconditioning matrix can be updated to better estimate the noise standard deviation using \mbox{$\matr{P^{k+1}}=\mathrm{diag}(\sqrt{\matr{P^k}\vect{y^k}})$}.

\section{Experimental Results}
\label{sec:Experiments}

\subsection{ADMM Parameters Setting}
The main hyper-parameters of the ADMM are the number of iterations $N$, the initial penalty parameter $\rho^0$, and it's increase factor $\alpha$ in Eq.~\eqref{eq:rho-decrease}.
The additional parameter $\sigma$ in Eq~\eqref{eq:PRLSQ} depends on the problem and corresponds to the Gaussian noise standard deviation in the measurement vector $\vect{b}$. For noise-free applications (i.e. completion, interpolation, noise-free demosaicing), using the theoretical value $\sigma=0$ would remove the regularization term. Therefore, in order to keep the benefit of the regularization in these applications, we use a small value $\sigma=1/255$. Also note that this parameter does not appear in the case of Poisson denoising in Eq.~\eqref{eq:PoissonDenPb} since the noise standard deviation only depends on the ground truth image.

\subsubsection{Completion, Interpolation and Demosaicing}

For the completion, interpolation and demosaicing (either with or without noise) the parameters setting is inspired from~\cite{zhang2021plug}: since the noise level of the denoiser is globally controlled at each iteration by $\sigma^k_{den}=\sigma/\sqrt{\rho^k}$, we choose $\rho^0$ such that $\sigma^0_{den}$ is sufficiently large to remove initialization noise or artifacts at the first iteration (regardless of the loss in details). For example, in the completion problem, we use zero-initialisation and a large value of $\sigma^0_{den}=1$. For the interpolation and demosaicing problems, more accurate initialisation can be performed, thus we use a smaller value of $\sigma^0_{den}=50/255$.
The parameter $\alpha$ is then determined so that $\sigma^k_{den}$ decreases down to $\sigma^N_{den}=\sigma$, in order to best preserve the details in the final image. Therefore, we have $\sigma^k_{den}=\sigma^N_{den}/\sqrt{\rho^k}$. Using this equation for $k=0$ and $k=N$, we can compute the parameters values:
\begin{equation}
\label{eq:params-setting}
\rho^0=\left(\frac{\sigma^N_{den}}{\sigma^0_{den}}\right)^2 \quad\text{and}\quad \alpha=\left(\frac{1}{\rho^0}\right)^{1/N}
\end{equation}
Note that the noise level assumed by the denoiser is also controlled locally by the preconditoning matrix $\matr{P}$ in Eq.\eqref{eq:y-update-preco}. However, our preconditioning strategy for these problems is such that $\matr{P}_{i,i}=1$ if $i$ is the index of a sampled pixel (see Section~\ref{ssec:CompleteAndInterp}). Hence at the last iteration, the image is denoised assuming the correct noise standard deviation $\sigma^N_{den}=\sigma$ for the sampled pixels. In our experiments, we can thus compare our preconditioned PnP-ADMM to the non-preconditioned version using the same parameterization of $\sigma^0_{den}$ and $\sigma^N_{den}$.

The preconditioning matrix definition in Eqs.~\eqref{eq:P_interp1} and~\eqref{eq:P_interp2} introduces the additional parameters $\sigma_f$, controlling the blurring of the mask at each iteration, and $\epsilon$, controlling the maximum preconditioning value $p_{max}=(1+\epsilon)/\epsilon$. In all the experiments, we use $p_{max}=10$ (i.e. $\epsilon=1/9$). The parameter $\sigma_f$ is set according to the number of iterations $N$ using \mbox{$\sigma_f=\sigma_f^{last}/\sqrt{N}$} as explained in Section~\ref{ssec:CompleteAndInterp}, where $\sigma_f^{last}$ is the blurring parameter at the last iteration that we set empirically for each application (see details in the respective sections and parameters summary in Table~\ref{tab:params}).

\subsubsection{Poisson Denoising}
\label{ssec:Params-Poisson}
For the problem of Poisson denoising, the noise level at each pixel and each iteration is fully controlled by the preconditioning matrix as explained in section~\ref{ssec:Poisson}, which removes the need for the penalty parameter. Therefore we simply use $\rho^0=1$ and $\alpha=1$.

Nevertheless in order to evaluate the advantage of the preconditioning, we also implement the non-preconditioned version by taking $\matr{P}=\matr{I}$ and using the $\netCst{}$ network. In this case, the noise standard deviation assumed by the denoiser in Eq.~\eqref{eq:y-update-preco} is $1/\sqrt{\rho^k}$ at each iteration $k$. Given an input image data in the range $[0,\lambda]$ and with Poisson noise, the highest possible noise standard deviation is $\sqrt{\lambda}$. Hence, we initialize $\rho_0$ so that $1/\sqrt{\rho^0}=\sqrt{\lambda}$. We found empirically that the best results are always obtained by decreasing the denoiser's standard deviation down to $1/\sqrt{\rho^N}=\sqrt{\lambda}/2$ at the last iteration $N$. Therefore in our experiments without preconditioning, we always use $\rho^0=1/\lambda$ and $\alpha=4^{1/N}$.

A summary of the parameters setting can be found in Table~\ref{tab:params}. It does not include the number of iterations $N$ which is studied more in detail for each application in the following sections.

\begin{table}
\setlength\tabcolsep{3pt}
\caption{Summary of the parameters setting. For the demosaicing with preconditioning, we use either $\sigma_f^{last}=0$ with the \netDem{} network or $\sigma_f^{last}=0.3$ with the \netVarCol{} network.}
\vspace{-10pt}
\renewcommand*{\arraystretch}{1.5}
\begin{tabular}{cc|cc|cccc}
\cline{3-8}
                                                     &            & \multicolumn{2}{c|}{without preco.} & \multicolumn{4}{c|}{with preco.}                                                         \\ \cline{3-8} 
                                                     &            & $\sigma^0_{den}$ & $\sigma^N_{den}$ &$\sigma^0_{den}$& $\sigma^N_{den}$ & $\sigma_f^{last}$ & \multicolumn{1}{c|}{$p_{max}$} \\ \hline
\multicolumn{2}{|c|}{completion}                                  & 1                & $\frac{1}{255}$  & 1              & $\frac{1}{255}$  & 0                 & \multicolumn{1}{c|}{10}     \\ \hline
\multicolumn{2}{|c|}{interpolation}                               & $\frac{50}{255}$ & $\frac{1}{255}$  &$\frac{50}{255}$& $\frac{1}{255}$  & 0.4               & \multicolumn{1}{c|}{10}     \\ \hline
\multicolumn{1}{|c|}{\multirow{2}{*}{demosaicing}}   & noise-free & $\frac{50}{255}$ & $\frac{1}{255}$  &$\frac{50}{255}$& $\frac{1}{255}$  & 0 or 0.3          & \multicolumn{1}{c|}{10}     \\
\multicolumn{1}{|c|}{}                           & noise $\sigma$ & $\frac{50}{255}$ & $\sigma$         &$\frac{50}{255}$& $\sigma$         & 0 or 0.3          & \multicolumn{1}{c|}{10}     \\ \hline
\\[-8pt] \cline{3-6}
&                                                                 & \multicolumn{2}{c|}{without preco.} & \multicolumn{2}{c|}{with preco.}  &                   & \\ \cline{1-6}
\multicolumn{2}{|c|}{\multirow{2}{*}{\specialcell{Poisson denoising\\(peak $\lambda$)}}}
                                                                  & $\rho^0$            & $\alpha$      & $\rho^0$ & \multicolumn{1}{c|}{$\alpha$} &            & \\ \cline{3-6}
\multicolumn{2}{|c|}{}                                            & $\frac{1}{\lambda}$ & $4^{1/N}$     & 1        & \multicolumn{1}{c|}{1}        &            & \\ \cline{1-6}
\end{tabular}
\label{tab:params}
\vspace{-10pt}
\end{table}

\subsection{Interpolation}
\label{ssec:res-interp}

\begin{figure}[t]
\centering
\begin{minipage}[]{.32\linewidth}
	\centerline{\includegraphics[width=\linewidth]{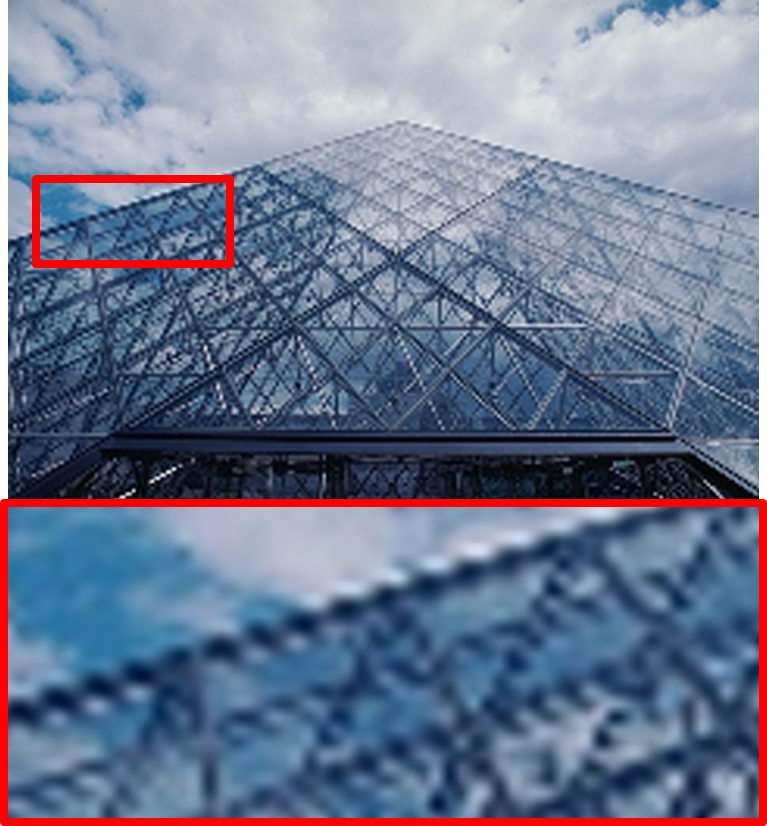}}
	\centerline{\small{(a) Bicubic}}
\end{minipage}
\begin{minipage}[h]{.32\linewidth}
	\centerline{\includegraphics[width=\linewidth]{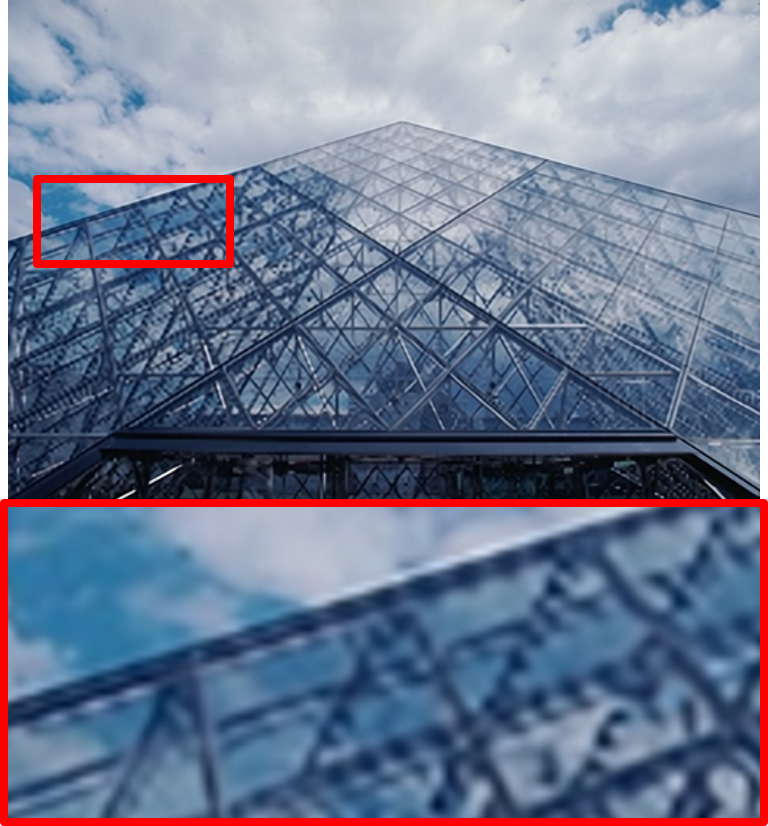}}
	\centerline{\small{(b) Ours: $\sigma_f^{last}=0$}}
\end{minipage}
\begin{minipage}[h]{.32\linewidth}
	\centerline{\includegraphics[width=\linewidth]{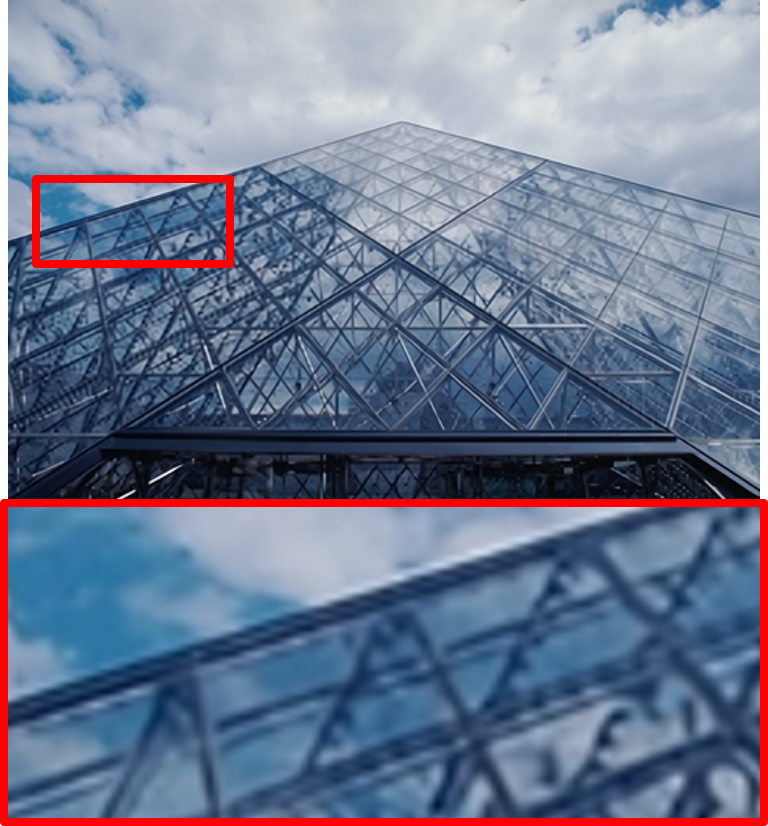}}
	\centerline{\small{(c) Ours: $\sigma_f^{last}=0.4$}}
\end{minipage}
\vspace{-5pt}
\caption{Interpolation for x2 upsampling using either: (a) bicubic interpolation, (b) our method without $\matr{P}$ update (i.e. $\sigma_f^{last}=0$), (c) our method with $\sigma_f^{last}=0.4$. For (b) and (c), we used $N=6$ iterations.}
\label{fig:preco_blur}
\vspace{-3pt}
\end{figure}

\begin{figure}[t]
\centering
\begin{minipage}[h]{.49\linewidth}
	\centerline{\includegraphics[width=\linewidth]{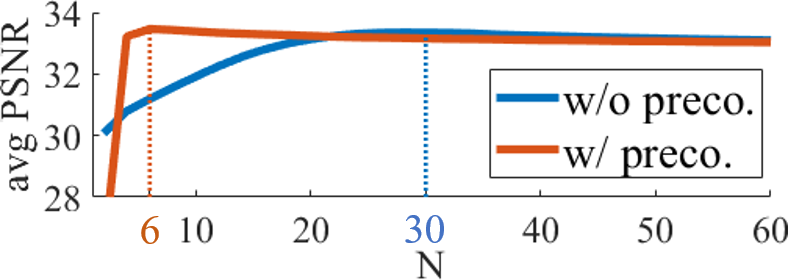}}
	\vspace{-3pt}
	\centerline{\small{(a) x2 interpolation.}}
\end{minipage}
\vspace{5pt}
\begin{minipage}[h]{.49\linewidth}
	\centerline{\includegraphics[width=\linewidth]{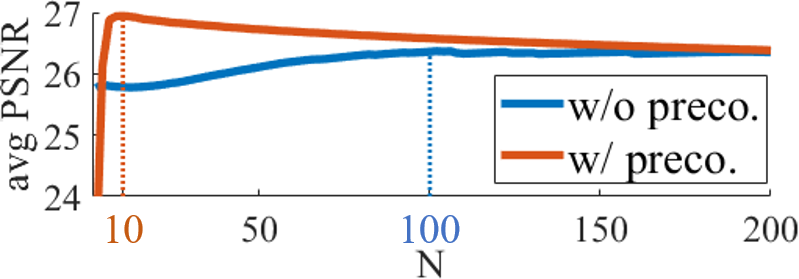}}
	\vspace{-3pt}
	\centerline{\small{(b) x4 interpolation}}
\end{minipage}
\vspace{5pt}
\begin{minipage}[h]{.49\linewidth}
	\centerline{\includegraphics[width=\linewidth]{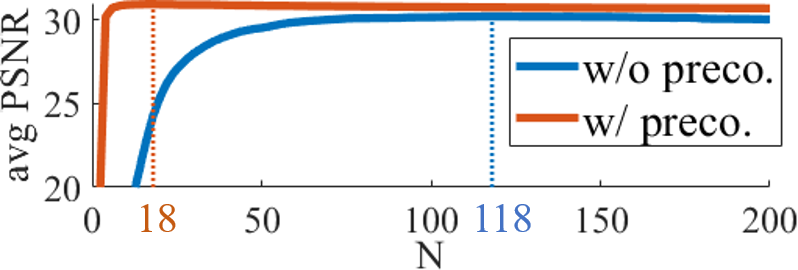}}
	\vspace{-3pt}
	\centerline{\small{(c) completion (20\%)}}
\end{minipage}
\begin{minipage}[h]{.49\linewidth}
	\centerline{\includegraphics[width=\linewidth]{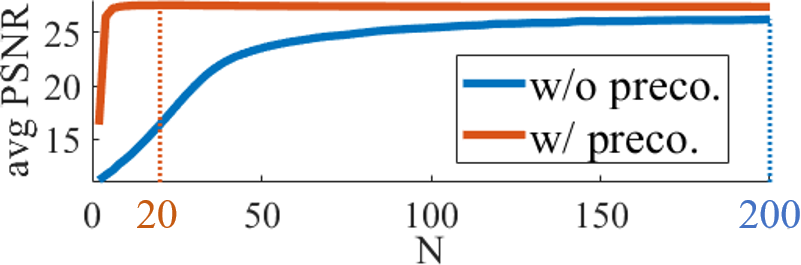}}
	\vspace{-3pt}
	\centerline{\small{(d) completion (10\%)}}
\end{minipage}
\vspace{-9pt}
\caption{Results of the PnP-ADMM either with or without preconditioning with respect to the number of iterations $N$ for: (a) x2 interpolation, (b) x4 interpolation, (c) completion from 20\% pixels, (d) completion from 10\% pixels. The plots show the average PSNR over the Set5 Dataset.}
\vspace{-3pt}
\label{fig:N_interp_complete}
\end{figure}

\begin{figure}[t!]
\centering
\begin{minipage}[h]{.49\linewidth}
	\centerline{\includegraphics[width=\linewidth]{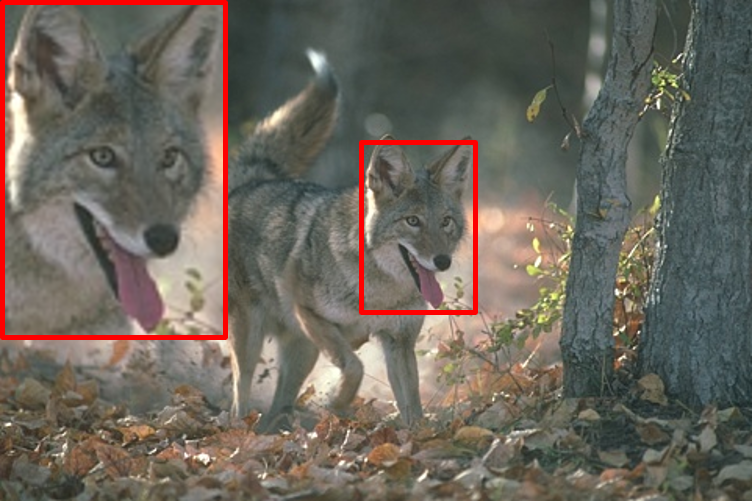}}
	\vspace{-3pt}
	\centerline{\small{(a) Ground Truth}}
	\centerline{\small{}}
\end{minipage}
\vspace{3pt}
\begin{minipage}[h]{.49\linewidth}
	\centerline{\includegraphics[width=\linewidth]{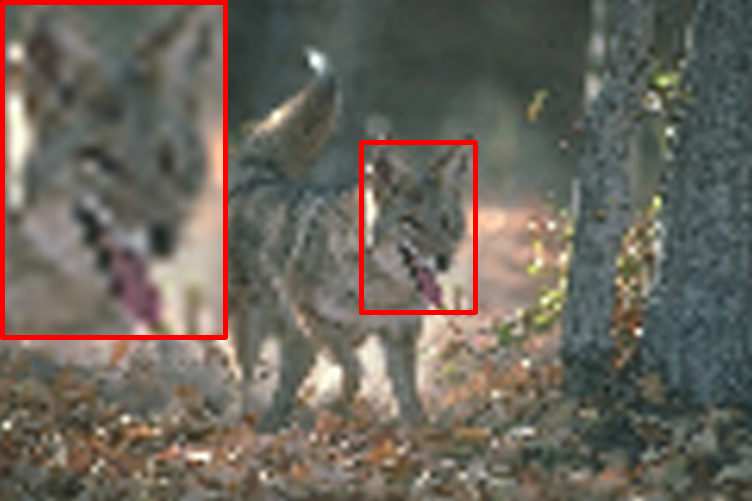}}
	\vspace{-3pt}
	\centerline{\small{(b) bicubic interpolation}}
	\centerline{\small{PSNR=25.30}}
\end{minipage}
\vspace{3pt}
\begin{minipage}[h]{.49\linewidth}
	\centerline{\includegraphics[width=\linewidth]{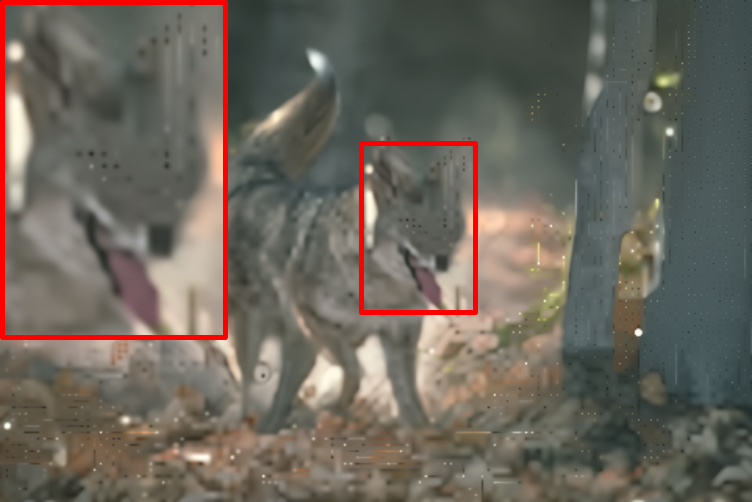}}
	\vspace{-3pt}
	\centerline{\small{(c) No Preconditioning N=100}}
	\centerline{\small{PSNR=25.37}}
\end{minipage}
\begin{minipage}[h]{.49\linewidth}
	\centerline{\includegraphics[width=\linewidth]{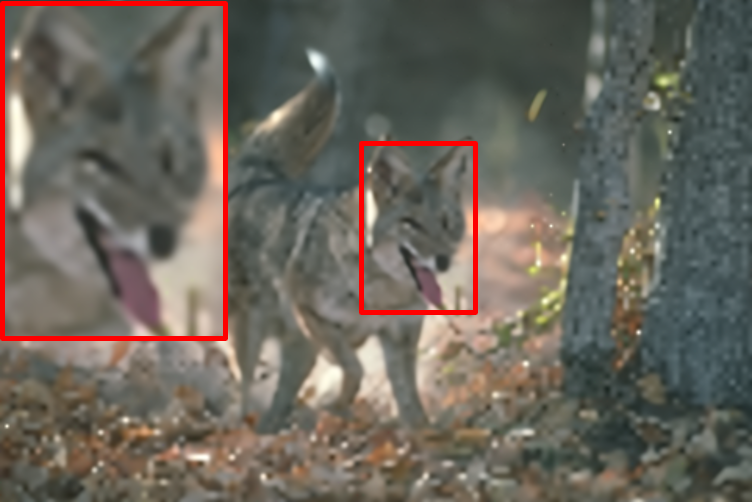}}
	\vspace{-3pt}
	\centerline{\small{(d) With Preconditioning N=10}}
	\centerline{\small{PSNR=25.84}}
\end{minipage}
\vspace{-8pt}
\caption{Example of x4 interpolation results.}
\label{fig:interp-x4}
\vspace{-8pt}
\end{figure}

\begin{table}[t!]
\centering
\caption{Interpolation results for x2 and x4 factors (average PSNR over each dataset). In each case, the number of iterations $N$ giving the best results is used for the PnP-ADMM.}
\vspace{-8pt}
\begin{tabular}{ll|cc|}
\cline{3-4}
                                          &                               & Set5  & CBSD68 \\ \hline
\multicolumn{1}{|l|}{\multirow{3}{*}{x2}} & bicubic interpolation         & 31.63 & 27.85 \\
\multicolumn{1}{|l|}{}                    & PnP-ADMM w/o preco. ($N=30$)  & 33.34 & 28.85 \\
\multicolumn{1}{|l|}{}                    & PnP-ADMM w/ preco. ($N=6$)    & 33.47 & 29.00 \\ \hline
\multicolumn{1}{|l|}{\multirow{3}{*}{x4}} & bicubic interpolation         & 25.66 & 23.20 \\
\multicolumn{1}{|l|}{}                    & PnP-ADMM w/o preco. ($N=100$) & \multicolumn{1}{l}{26.36} & 23.65 \\
\multicolumn{1}{|l|}{}                    & PnP-ADMM w/ preco. ($N=10$)   & \multicolumn{1}{l}{26.94} & 23.90 \\ \hline
\end{tabular}
\label{tab:res-interp}
\vspace{-12pt}
\end{table}

\begin{figure*}[t]
\centering
\begin{minipage}[h]{.162\linewidth}
	\centerline{\includegraphics[width=\linewidth]{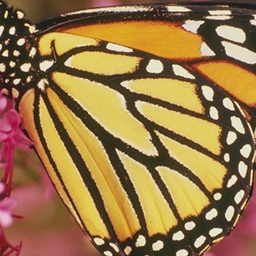}}
	\vspace{2pt}
	\centerline{\includegraphics[width=\linewidth]{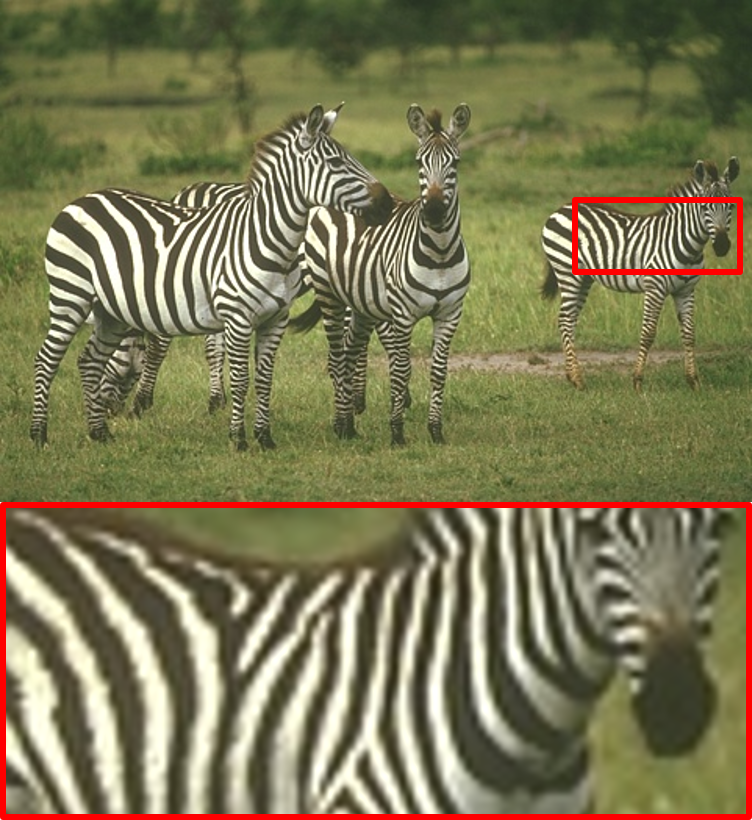}}
	\centerline{\footnotesize{(a) Ground Truth}}
\end{minipage}
\begin{minipage}[h]{.162\linewidth}
	\centerline{\includegraphics[width=\linewidth]{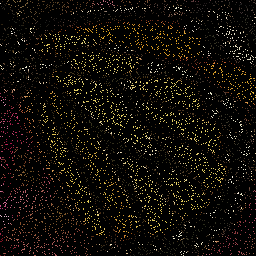}}
	\vspace{2pt}
	\centerline{\includegraphics[width=\linewidth]{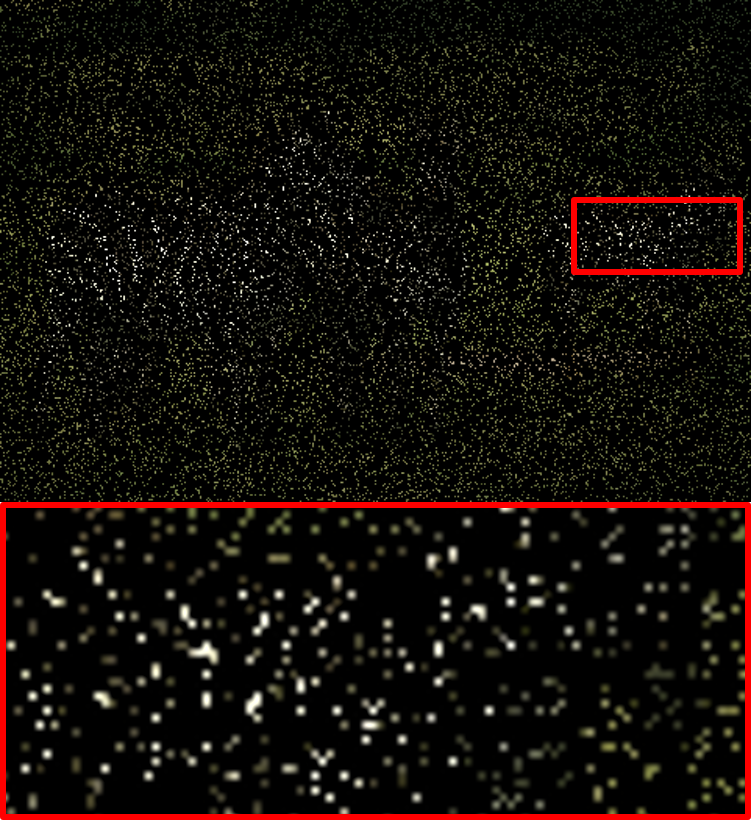}}
	\centerline{\footnotesize{(b) Input ($10\%$ pixels)}}
\end{minipage}
\begin{minipage}[h]{.162\linewidth}
	\centerline{\includegraphics[width=\linewidth]{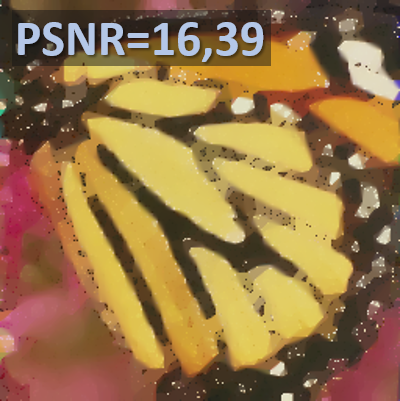}}
	\vspace{2pt}
	\centerline{\includegraphics[width=\linewidth]{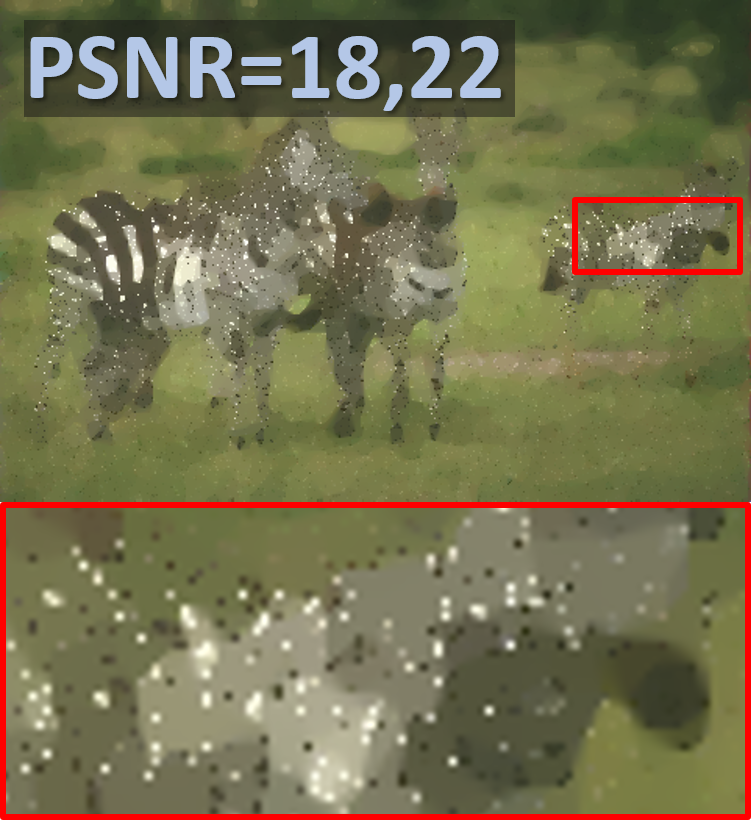}}
	\centerline{\footnotesize{(c) Total Variation}}
\end{minipage}
\begin{minipage}[h]{.162\linewidth}
	\centerline{\includegraphics[width=\linewidth]{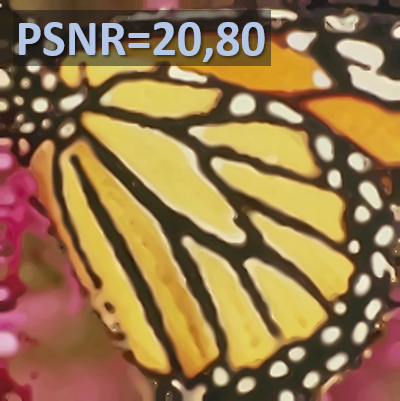}}
	\vspace{2pt}
	\centerline{\includegraphics[width=\linewidth]{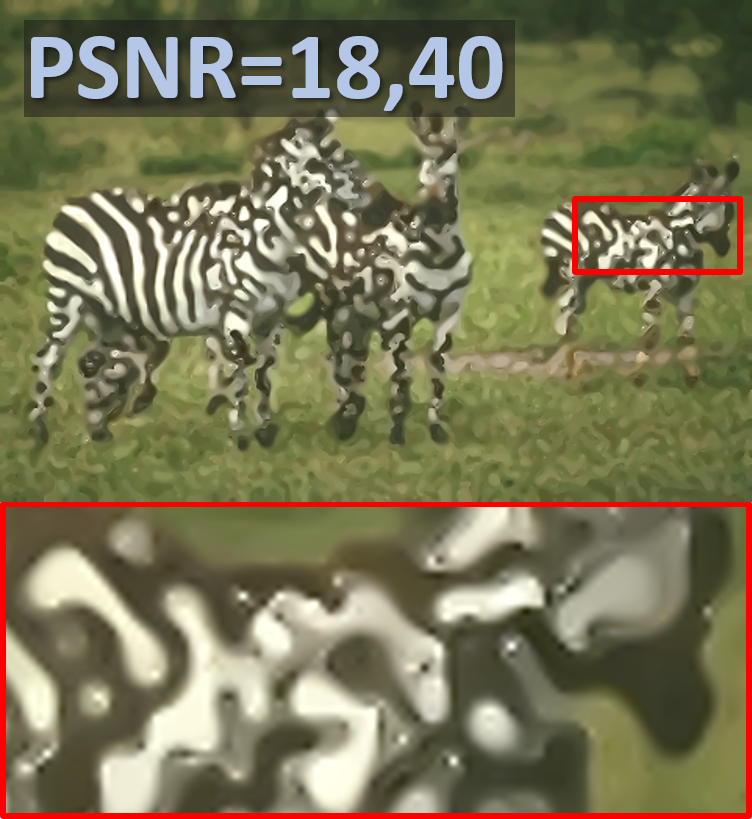}}
	\centerline{\footnotesize (d) Diffusion (EED)}
\end{minipage}
\begin{minipage}[h]{.162\linewidth}
	\centerline{\includegraphics[width=\linewidth]{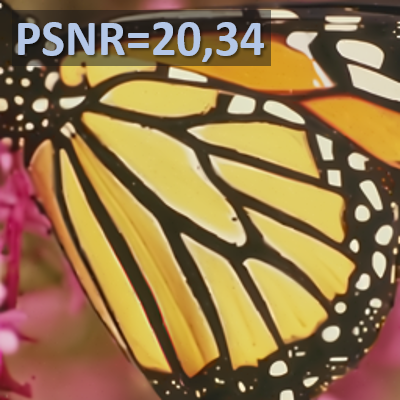}}
	\vspace{2pt}
	\centerline{\includegraphics[width=\linewidth]{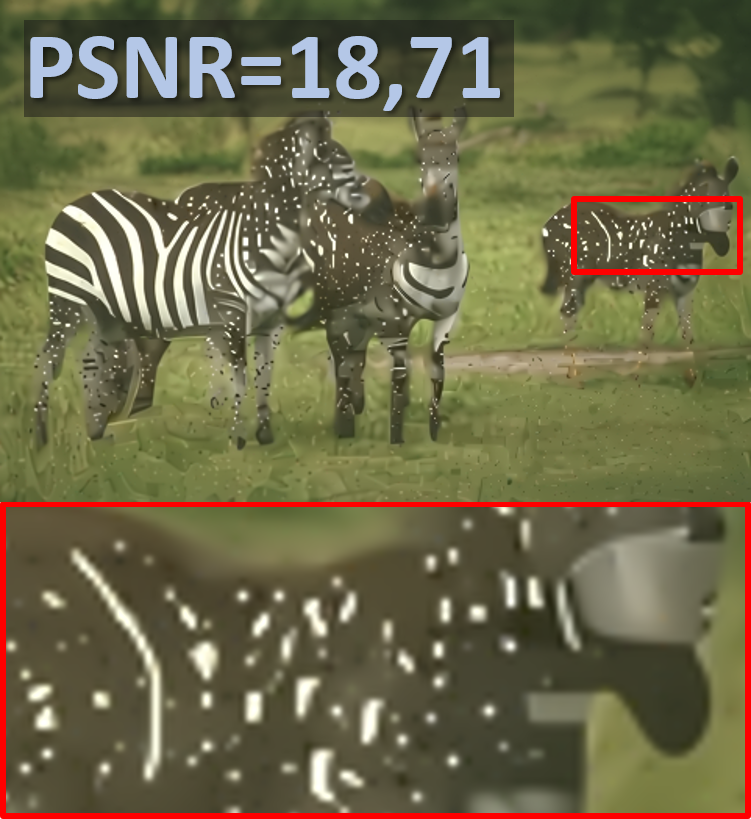}}
	\centerline{\footnotesize{(e) PnP-ADMM}}
\end{minipage}
\begin{minipage}[h]{.162\linewidth}
	\centerline{\includegraphics[width=\linewidth]{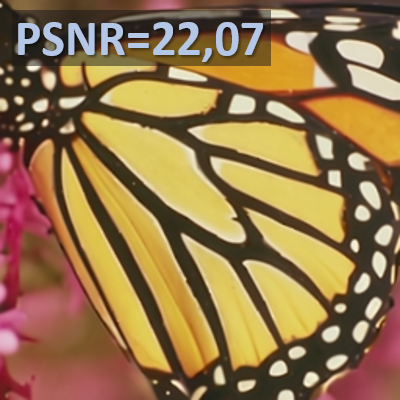}}
	\vspace{2pt}
	\centerline{\includegraphics[width=\linewidth]{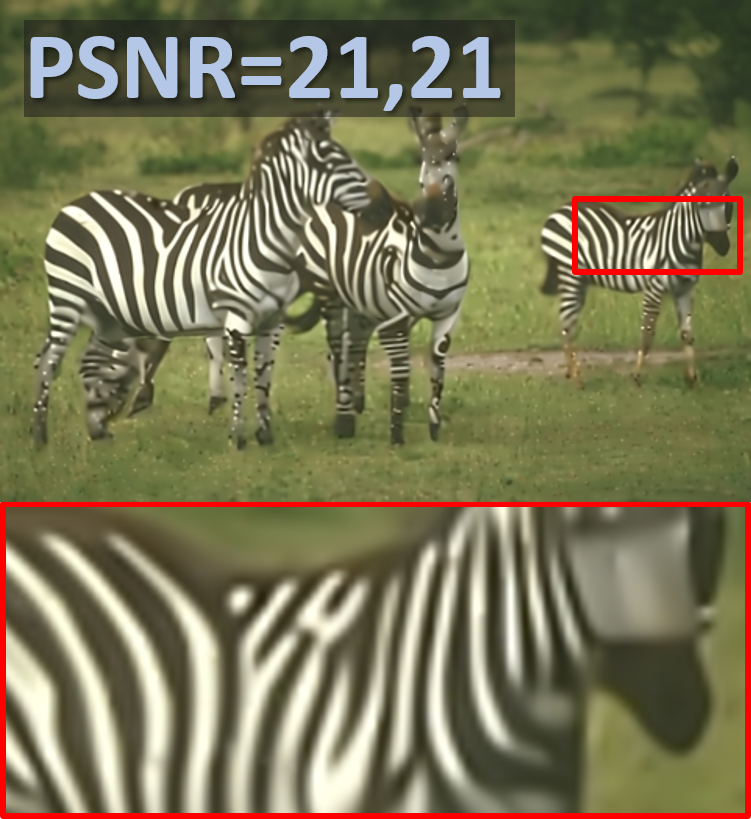}}
	\centerline{\footnotesize{(f) PnP-ADMM (preco.)}}
\end{minipage}
\vspace{-10pt}
\caption{Results of completion from $10\%$ pixels for the butterfly image using (c) Total Variation regularization, (d) Edge-Enhancing Diffusion, (e) PnP-ADMM without preconditioning (using $N=200$ iterations), (f) PnP-ADMM with our preconditioning  (using $N=20$ iterations).}
\label{fig:res-complete}
\vspace{-5pt}
\end{figure*}

The interpolation problem can be seen as a particular case of super-resolution, where the low resolution input image was generated without applying an antialiasing filter before sub-sampling the ground truth image pixels. As a result, the main difficulty of the interpolation problem is to remove aliasing effects.
First, we show in Fig.~\ref{fig:preco_blur} that our method better removes aliasing when we update the preconditioning matrix $\matr{P}$ by blurring the mask of sampled pixels, as described in Section~\ref{ssec:CompleteAndInterp} (see Eqs.~\eqref{eq:P_interp1}-\eqref{eq:P_interp2}). In this example, the results of the bicubic interpolation in Fig.~\ref{fig:preco_blur}(a) displays strong aliasing artifacts. In Fig.~\ref{fig:preco_blur}(b), using our preconditioned PnP-ADMM with $\sigma_f^{last}=0$ (i.e. no update of $\matr{P}$ and no blurring of the mask) partially removes the aliasing. However, better results are obtained in Fig.~\ref{fig:preco_blur}(c), when updating $\matr{P}$ using $\sigma_f^{last}=0.4$.

Although many super-resolution algorithms exist and excellent performances have been obtained by deep learning techniques, these methods are generally trained assuming that the degradation already includes a given antialiasing filter. Therefore these methods are not suitable for the interpolation. Hence, we only compare our approach to the non-preconditioned PnP-ADMM, as well as the bicubic interpolation which we also use as an initialisation in the ADMM.
The results of the PnP-ADMM either with or without preconditioning depends on the chosen number of iterations $N$. Fig. \ref{fig:N_interp_complete}(a)-(b) shows that without preconditioning, the best results are obtained using respectively $N=30$ and $N=100$ iterations for x2 and x4 interpolation, while with our preconditioning, we only need to use $N=6$ and $N=10$ respectively. Using these values for the parameter $N$ in each case, the results in Table~\ref{tab:res-interp} and Fig.~\ref{fig:interp-x4} show that in addition to the faster convergence, our preconditioning improves the final results of the PnP-ADMM both in x2 and x4 interpolation.

\subsection{Completion}
\label{ssec:res-completion}

\begin{table}
\centering
\vspace{-3pt}
\caption{Completion results for $20\%$ and $10\%$ rates (average PSNR over each dataset). In each case, the number of iterations $N$ giving the best results is used for the PnP-ADMM.}
\vspace{-10pt}
\begin{tabular}{ll|cc|}
\cline{3-4}
                                          &                               & Set5  & CBSD68 \\ \hline
\multicolumn{1}{|l|}{\multirow{4}{*}{$20\%$}} & Total Variation regularization & 26.11 & 24.80 \\
\multicolumn{1}{|l|}{}                    & Edge Enhancing Diffusion           & 28.61 & 25.55 \\
\multicolumn{1}{|l|}{}                    & PnP-ADMM w/o preco. ($N=118$)      & 30.20 & 26.75 \\
\multicolumn{1}{|l|}{}                    & PnP-ADMM w/ preco. ($N=18$)        & 30.94 & 27.43 \\ \hline
\multicolumn{1}{|l|}{\multirow{4}{*}{$10\%$}} & Total Variation regularization & 22.86 & 22.66 \\
\multicolumn{1}{|l|}{}                    & Edge Enhancing Diffusion      & \multicolumn{1}{l}{25.85} & 23.43 \\
\multicolumn{1}{|l|}{}                    & PnP-ADMM w/o preco. ($N=200$) & \multicolumn{1}{l}{26.20} & 24.06 \\
\multicolumn{1}{|l|}{}                    & PnP-ADMM w/ preco. ($N=20$)   & \multicolumn{1}{l}{27.52} & 24.95 \\ \hline
\end{tabular}
\label{tab:res-complete}
\vspace{-5pt}
\end{table}

\begin{table*}
\centering
\caption{Demosaicing results (average PSNR over each dataset) for the Bayer pattern in the noise-free scenario ($\sigma=0$). For our preconditioned PnP-ADMM method, we show the results using either the generic denoiser \netVarCol{} or the specialized denoiser for demosaicing \netDem{}. The best and second best results are in \red{red} and \blue{blue} respectively.}
\vspace{-10pt}
\begin{tabular}{l|c|c|c|c|c|c|c|}
\cline{2-8}
&
\multirow{3}{*}{\specialcell{Malvar et al.\\ (Matlab) \cite{Malvar04}}}&
\multirow{3}{*}{\specialcell{Gharbi et al.\\ (DeepJoint)\cite{gharbi2016deep}}}&
\multirow{3}{*}{\specialcell{Henz et al.\\\cite{henz2018deep}}}&
\multirow{3}{*}{\specialcell{Kokkinos et al.\\ (MMNet)~\cite{kokkinos2019}}}&
\multicolumn{3}{c|}{PnP-ADMM}  \\ \cline{6-8} 
&   &   &   &   & no preconditioning & \multicolumn{2}{c|}{with preconditioning} \\ \cline{7-8} 
&   &   &   &   & (\netCst{})        & \netVarCol{}         &         \netDem{}  \\ \hline
\multicolumn{1}{|l|}{Kodak}    & 34.68 & 41.85 & 42.05       & 40.26 & 41.45 & \red{42.37} & \blue{42.32} \\
\multicolumn{1}{|l|}{McMaster} & 34.46 & 39.13 & \red{39.50} & 36.84 & 39.24 & 39.27       & \blue{39.31} \\ \hline
\end{tabular}
\vspace{-5pt}
\label{tab:demosaick}
\end{table*}

\begin{figure*}
\centerline{\includegraphics[width=\linewidth]{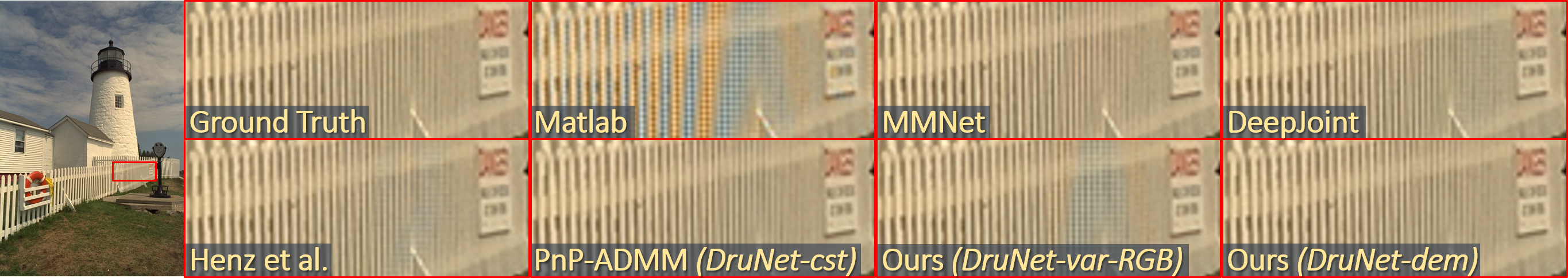}}
\vspace{-12pt}
\caption{Demosaicing results in the noise-free scenario. The ``PnP-ADMM (\netCst{})'' result corresponds to the non-preconditioned version, while our method is preconditioned and uses either the generic denoiser \netVarCol{} or the specialized one \netDem{}.}
\vspace{-2pt}
\label{fig:demosaick_res}
\end{figure*}

\begin{table*}
\centering
\caption{Demosaicing results (average PSNR over each dataset) for the Bayer pattern in the noisy scenario with noise of standard deviation $\sigma=10/255$ or $\sigma=20/255$. For our preconditioned PnP-ADMM method, we show the results using either the generic denoiser \netVarCol{} or the specialized denoiser for demosaicing \netDem{}. The best and second best results are in \red{red} and \blue{blue} respectively.}
\vspace{-10pt}
\begin{tabular}{rl|c|c|c|c|c|c|c|}
\cline{3-8}
&   &
\multirow{3}{*}{\specialcell{Malvar et al.\\ (Matlab) \cite{Malvar04}}}&
\multirow{3}{*}{\specialcell{Gharbi et al.\\ (DeepJoint)~\cite{gharbi2016deep}}}&
\multirow{3}{*}{\specialcell{Kokkinos et al.\\ (MMNet)~\cite{kokkinos2019}}}&
\multicolumn{3}{c|}{PnP-ADMM}  \\ \cline{6-8} 
&   &   &   &   & no preconditioning & \multicolumn{2}{c|}{with preconditioning} \\ \cline{7-8} 
&   &   &   &   & (\netCst{})        & \netVarCol{}         &         \netDem{}  \\ \cline{2-8}
\multicolumn{1}{r|}{\multirow{2}{*}{$\sigma=10/255$}} & Kodak & 27.54  & 33.27 & 30.96 & 33.24 & \blue{33.90}  & \red{34.18} \\
\multicolumn{1}{r|}{}                              & McMaster & 27.69  & 33.18 & 30.19 & 33.35 & \blue{33.89}  & \red{34.06} \\\cline{2-8}
\multicolumn{1}{r|}{\multirow{2}{*}{$\sigma=20/255$}} & Kodak & 22.38  & 30.04 & 23.81 & 29.92 & \blue{30.84}  & \red{31.18} \\
\multicolumn{1}{r|}{}                              & McMaster & 22.75  & 30.18 & 23.65 & 30.09 & \blue{31.09}  & \red{31.36} \\ \cline{2-8}
\end{tabular}
\label{tab:demosaick_denoise}
\vspace{-8pt}
\end{table*}


For the completion problem, we evaluate the results using a rate of either $20\%$ or $10\%$ of known pixels. Unlike the interpolation problem, the known pixels do not form a regular grid and are instead selected randomly, which prevents the aliasing artifacts.
For this reason, contrary to what we observed for the interpolation in Fig.~\ref{fig:preco_blur}, updating the preconditioning matrix $\matr{P}$ with a blurred version of the mask does not improve the completion results. Hence, we use $\sigma_f^{last}=0$ in the following experiments. 

Based on the results in Fig.~\ref{fig:N_interp_complete}(c)-(d), we use $N=18$ and $N=20$ iterations respectively for the $20\%$ and $10\%$ completion problems when using our preconditioning. Similarly to the interpolation problem, the convergence of the non-preconditioned PnP-ADMM is significantly slower: the best results for the Set5 dataset are obtained using respectively $N=118$ and $N=200$ for $20\%$ and $10\%$ completion.

For further comparisons, we also provide completion results obtained with the conventional total variation (TV) regularization, using the implementation in \cite{dahl2009}. We also compare our results to a state of the art diffusion based method. A review of diffusion methods in \cite{PDE-EED} have shown that the best completion performance was obtained with a non-linear anisotropic diffusion scheme referred to as edge enhancing diffusion (EED). The EED was also exploited within compression schemes in more recent works (e.g. \cite{andris2016},\cite{schmaltz14}), where only a small percentage of pixels are encoded and the remaining ones must be recovered by the decoder. Therefore, we use the EED scheme for our comparisons.

The results in Fig.~\ref{fig:res-complete} show that, while the EED diffusion better recovers large image structures than solving the inverse problem with TV regularizatioin, the PnP-ADMM scheme retrieves finer details and sharper edges thanks to the advanced regularization provided by the trained denoiser. However, without our preconditioning, some important structures of the image may be lost as shown in Fig.~\ref{fig:res-complete}(e) (e.g. white spots on the butterfly, zebras' stripes).
On the other hand, our preconditioned version recovers all the structures that could be inferred from the input data, hence resulting in a large gain in PSNR. Note that this observation is consistent with the interpolation results in Fig.~\ref{fig:interp-x4}, where important details of the image are lost in the non-preconditioned PnP-ADMM but are preserved with our preconditioning.
The average PSNR results in Table~\ref{tab:res-complete} confirm the superiority of the proposed preconditioned PnP-ADMM over the other approaches for image completion from either $10\%$ or $20\%$ pixels.

\subsection{Demosaicing}
\label{ssec:resDem}

\begin{figure}
\centering
\begin{minipage}[h]{.49\linewidth}
	\centerline{\includegraphics[width=\linewidth]{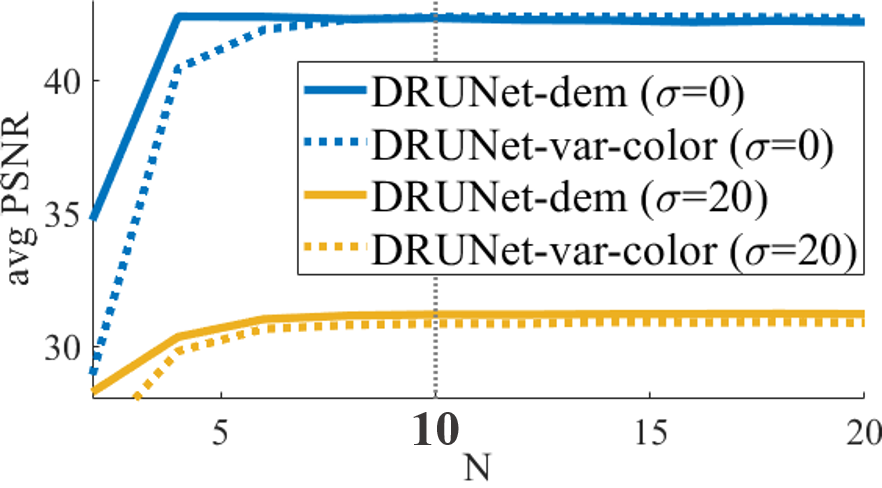}}
	\vspace{-3pt}
	\centerline{\small{(a) with preconditioning.}}
\vspace{4pt}
\end{minipage}
\begin{minipage}[h]{.49\linewidth}
	\centerline{\includegraphics[width=\linewidth]{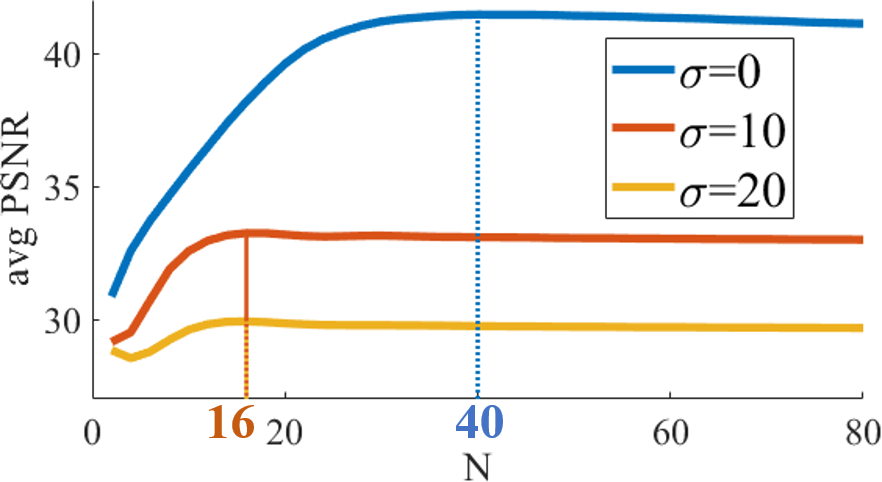}}
	\vspace{-3pt}
	\centerline{\small{(b) without preconditioning}}
\end{minipage}
\vspace{-6pt}
\caption{Results of the PnP-ADMM for demosaicing with respect to the number of iterations $N$: (a) with preconditioning, (b) without preconditioning. The plots show the average PSNR over the Kodak Dataset.}
\vspace{-3pt}
\label{fig:N_demosaick}
\end{figure}

For the demosaicing, we evaluate the results both in the noise-free scenario and in the presence of Gaussian noise of standard deviation $\sigma=10$ or $\sigma=20$.
In each case, the input images are filtered by the traditional Bayer CFA, i.e. only one of the red green or blue components is known at each pixel.
The Bayer CFA forms a repetitive pattern of 2x2 blocks, each containing one red, one blue and two green pixels.
For our preconditioned PnP-ADMM method, we evaluate the results using either the generic denoising network \netVarCol{} or the specialized network \netDem{} that was trained for noise level patterns generated from the Bayer CFA, as explained in Section~\ref{ssec:demosaicing}.

\begin{figure}[t!]
\centerline{\includegraphics[width=\linewidth]{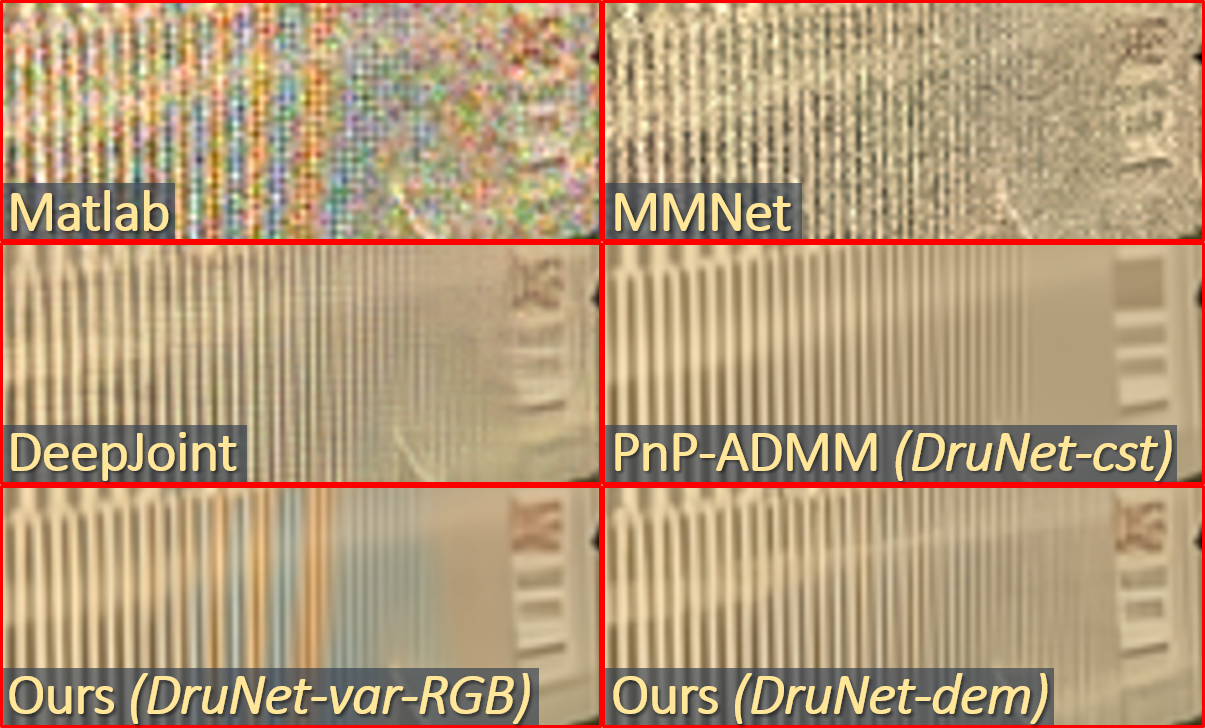}}
\vspace{-10pt}
\caption{Demosaicing results for the same detail as in Fig.~\ref{fig:demosaick_res}, but in the noisy scenario with Gaussian noise of standard deviation $\sigma=20$.}
\vspace{-10pt}
\label{fig:demosaick_denoise_res}
\end{figure}

\begin{figure*}[t]
\centering
\begin{minipage}[h]{.16\linewidth}
	\centerline{\includegraphics[width=\linewidth]{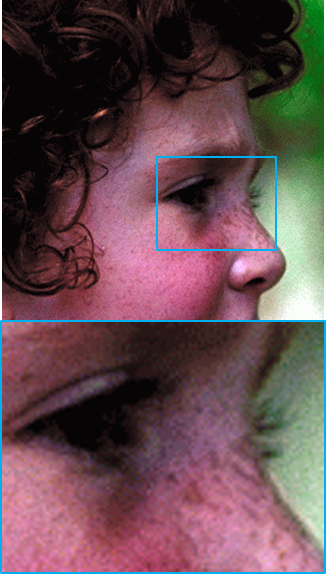}}
	\centerline{\small{(a) Ground Truth}}
	\centerline{\small{}}
\end{minipage}
\begin{minipage}[h]{.16\linewidth}
	\centerline{\includegraphics[width=\linewidth]{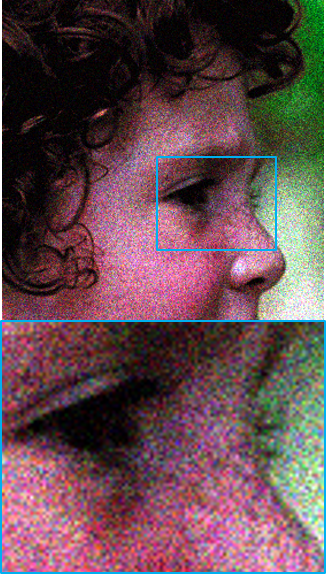}}
	\centerline{\small{(b) Noisy}}
	\centerline{\small{$\lambda=255/8$}}
	\vspace{-2.5pt}
\end{minipage}
\begin{minipage}[h]{.16\linewidth}
	\centerline{\includegraphics[width=\linewidth]{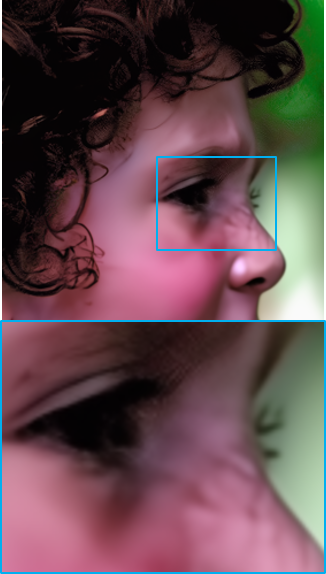}}
	\centerline{\small{(c) No Preco. N=6}}
	\centerline{\small{PSNR=30.15}}
\end{minipage}
\begin{minipage}[h]{.16\linewidth}
	\centerline{\includegraphics[width=\linewidth]{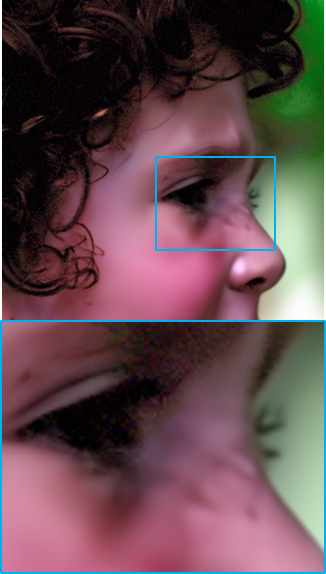}}
	\centerline{\small{(d) No Preco. N=100}}
	\centerline{\small{PSNR=29.49}}
\end{minipage}
\begin{minipage}[h]{.16\linewidth}
	\centerline{\includegraphics[width=\linewidth]{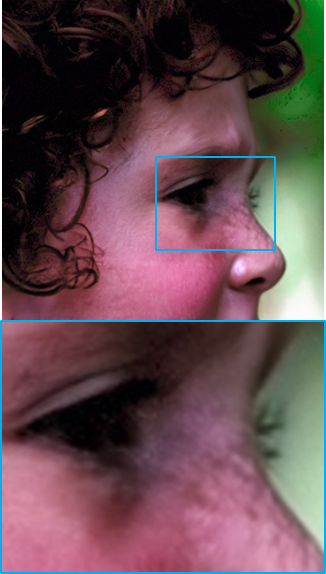}}
	\centerline{\small{(e) With Preco. N=6}}
	\centerline{\small{PSNR=30.28}}
\end{minipage}
\begin{minipage}[h]{.16\linewidth}
	\centerline{\includegraphics[width=\linewidth]{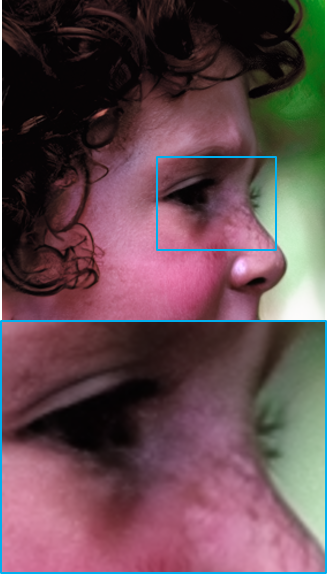}}
	\centerline{\small{(f) With Preco. N=100}}
	\centerline{\small{PSNR=30.76}}
\end{minipage}
\vspace{-1pt}
\caption{Poisson Denoising given a peak value $\lambda=255/8$. Results are shown for the PnP-ADMM either with preconditioning (e,f) or without (c,d) and using a number of iterations $N=6$ (c,e) or $N=100$ (d,f).}
\label{fig:den-Poisson-vizu}
\vspace{-5pt}
\end{figure*}

\begin{figure}
\centering
\begin{minipage}[h]{.9\linewidth}
	\centerline{\includegraphics[width=\linewidth]{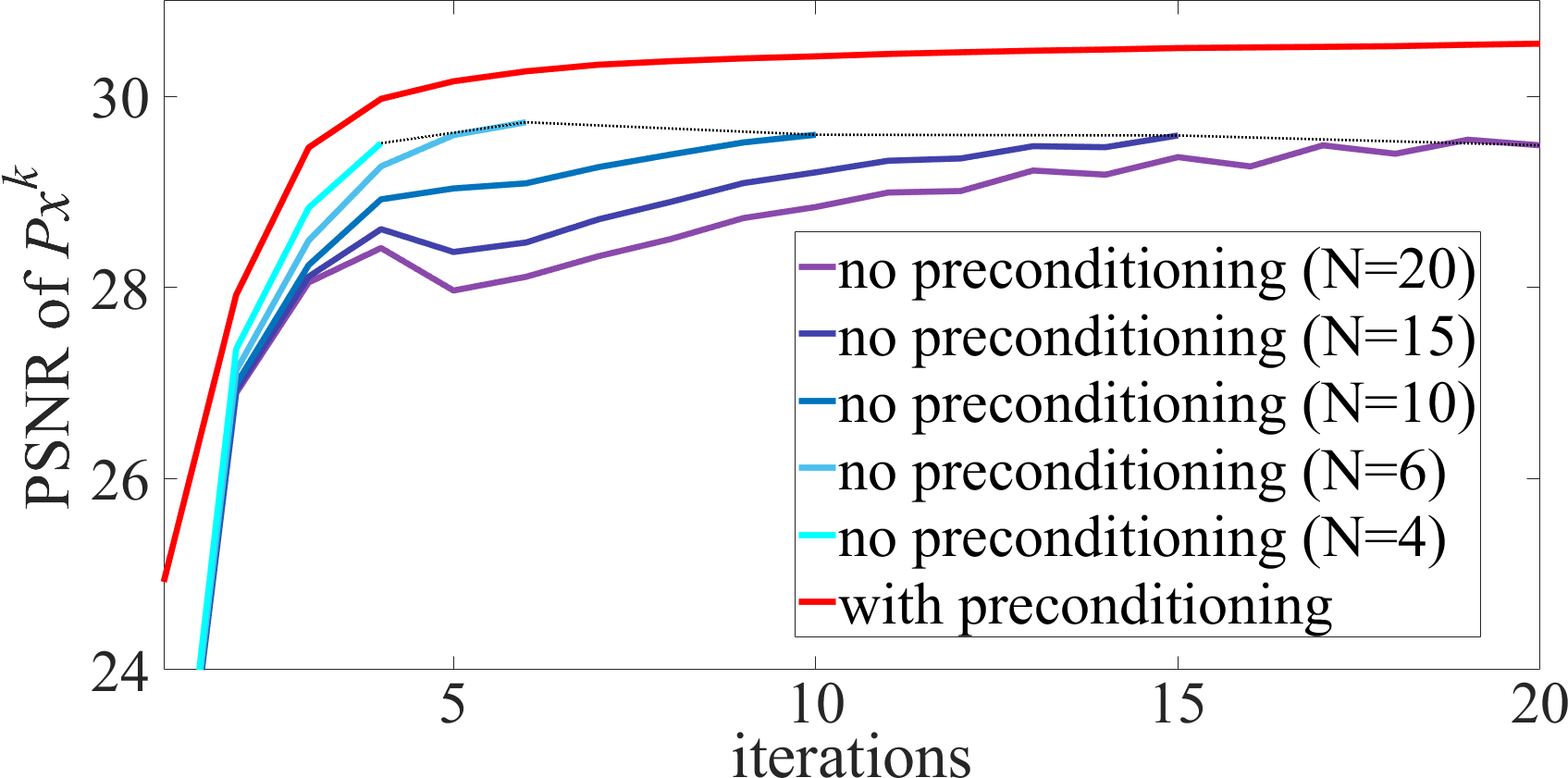}}
	\vspace{-3pt}
	\centerline{\small{(a) PSNR of the results at each iteration.}}
\vspace{4pt}
\end{minipage}
\begin{minipage}[h]{.9\linewidth}
	\centerline{\includegraphics[width=\linewidth]{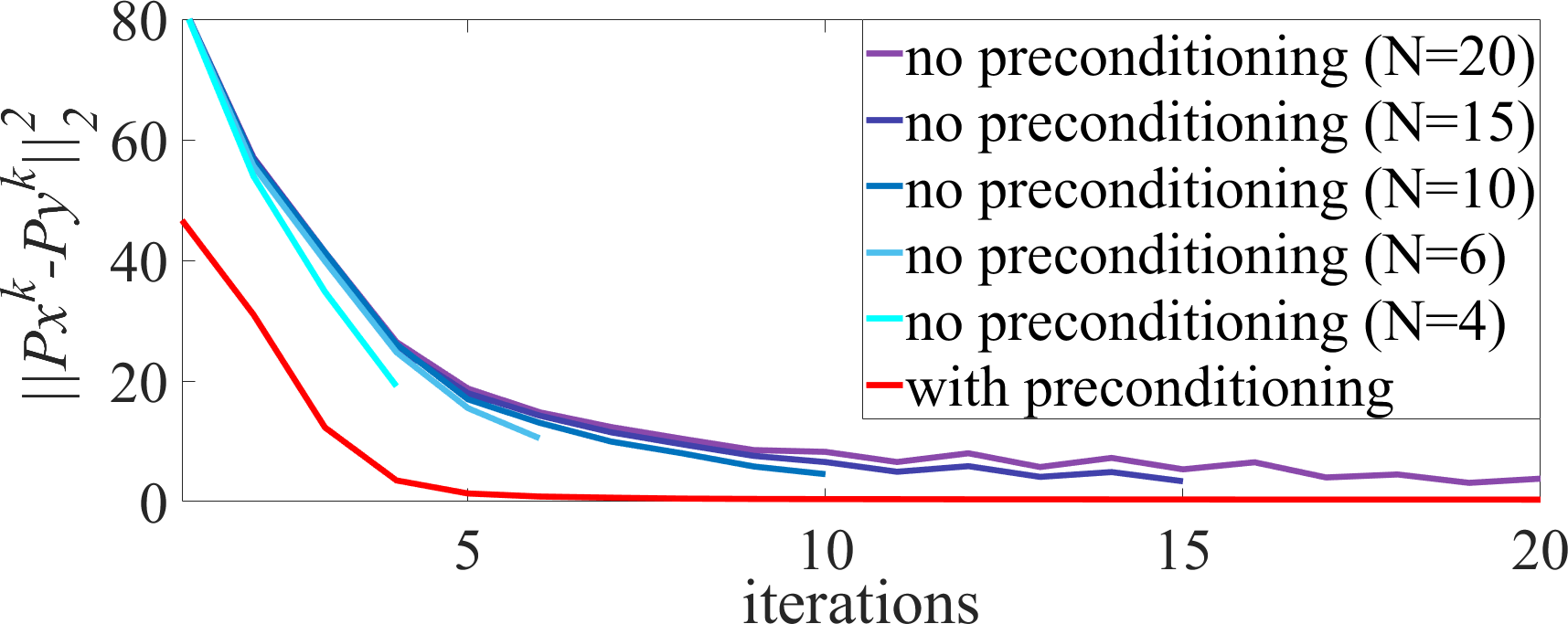}}
	\centerline{\small{(b) Mean Square Error between the two ADMM variables}}
\end{minipage}
\vspace{-2pt}
\caption{Convegence plots of the PnP-ADMM for Poisson denoising with $\lambda=255/8$ and for the image ``head'' (from Fig.~\ref{fig:den-Poisson-vizu}). For, the non-preconditioned case, several parameters depend on the fixed number of iterations $N$, thus, different curves are obtained with respect to $N$. With our preconditioning, the curves for different values of $N$ are superimposed on the red curve because no parameter depend on $N$.}
\vspace{-3pt}
\label{fig:den-Poisson-cv}
\end{figure}

Similarly to the interpolation problem, the repetition of the 2x2 block pattern in the Bayer CFA may cause artifacts comparable to aliasing. When using the generic network \netVarCol{} in our preconditioning scheme, these artifacts are better removed by updating the matrix $\matr{P}$ with a blurred version of the mask. In this case, we use $\sigma_f^{last}=0.3$ to control the blur at each iteration. However, when using the denoiser \netDem{} specialized for such patterns, the aliasing artifacts from the Bayer CFA are naturally removed without the need for altering the preconditioning matrix. Thus we use $\sigma_f^{last}=0$ in this case.

Note that for all the experiments, a fast initialisation is performed with Matlab's demosaicing method~\cite{Malvar04} that uses bilinear interpolation and gradient-based corrections.
Given this initialisation, we fix the number of iterations $N$ of the PnP-ADMM according to the preliminary experiment in Fig.~\ref{fig:N_demosaick}. Using our preconditioning (Fig.~\ref{fig:N_demosaick}(a)), close to optimal results are obtained using $N=10$, either with or without noise, and for both the \netVarCol{} and \netDem{} networks. More iterations are needed for the non-preconditioned PnP-ADMM which reaches its best performance using $N=40$ in the noise-free scenario ($\sigma=0$), and using $N=16$ in both noisy settings.

For the evaluation, we use the Kodak~\cite{kodak} and McMaster~\cite{McMaster} test datasets that are widely used for demosaicing benchmarks as they contain natural images which include challenging features for this application. We compare both versions of our preconditioned PnP-ADMM with the non-preconditioned version (based on the \netCst{} denoiser), as well as other reference methods including Matlab's demosaicing method by Malvar et al.~\cite{Malvar04} and the deep learning based methods of Gharbi et al. (DeepJoint)~\cite{gharbi2016deep}, Henz et al.~\cite{henz2018deep} and Kokkinos et al. (MMNet)~\cite{kokkinos2019}.

Table~\ref{tab:demosaick} and Fig.~\ref{fig:demosaick_res} present the results in the noise-free scenario. Our method performs among the best in average PSNR, indicating that it successfully recovers the fine details overall, and with greater accuracy than the non-preconditioned PnP-ADMM. However, in very challenging areas with high frequency patterns (see Fig.~\ref{fig:demosaick_res}), our preconditioning based on the generic network \netVarCol{} leaves some zipper and color artifacts. Note that, on closer inspection, similar artifacts also remain in the Henz et al. and DeepJoint methods. However, using the specialized denoiser \netDem{} in our scheme completely solves this issue. This shows that it was not a limitation of the preconditioning or the PnP-ADMM scheme itself, but rather a limitation of the \netVarCol{} denoiser that does not perform optimally when the noise level map is a structured pattern constructed from the Bayer CFA.

Similar conclusions can be drawn for the noisy scenario in Table~\ref{tab:demosaick_denoise} and Fig.~\ref{fig:demosaick_denoise_res}. Our preconditioned PnP-ADMM best recovers the image details, and using the specialized \netDem{} network prevents the color issues that may appear in challenging areas when using the generic \netVarCol{} network.

\subsection{Poisson Denoising}

For Poisson denoising, we compare the results of the PnP-ADMM either with or without our preconditioning. Note that the non-preconditioned version was previously described in~\cite{PoissonDen} for this application. However, the BM3D denoiser was used in ~\cite{PoissonDen}. For a fair evaluation of our approach, we use instead the \netCst{} network which gives the best performance when no preconditioning is required. For our preconditioned scheme, we use the \netVarCol{} denoiser since the Poisson noise variance depends not only on pixels' positions, but also on the colour component.

First, we analyse the results for moderate to high levels of Poisson noise (using $\lambda\in\{255$, $255/4$, $255/8\}$). Table~\ref{tab:den-Poisson} and Fig.~\ref{fig:den-Poisson-cv}(a) show that the the proposed preconditioning improves the results compared to the original PnP-ADMM, regardless of the fixed number of iterations $N$. Furthermore, a better convergence is obtained with our approach as shown in Fig.~\ref{fig:den-Poisson-cv}. While the results of the non-preconditioned PnP-ADMM start degrading when using more than $N=6$ iterations, our method successfully converges to the best result. The improved convergence is confirmed in Fig.~\ref{fig:den-Poisson-cv}(b) showing that the equality constraint between the two ADMM variables is better satisfied with our preconditioning.
The visual results in Fig.~\ref{fig:den-Poisson-vizu} also show that our approach can recover finer details. Using a large number of iterations further improves the PSNR in this case, although the visual results are similar (see Fig.~\ref{fig:den-Poisson-vizu}(e,f)). On the other hand, in the non-preconditioned PnP-ADMM, using a too large number of iterations visually degrades the results by removing details in the bright regions, while leaving more noise in the dark regions (see Fig.~\ref{fig:den-Poisson-vizu}(c,d)).

\begin{figure}[t]
\centering
\begin{minipage}[h]{.49\linewidth}
	\centerline{\includegraphics[width=\linewidth]{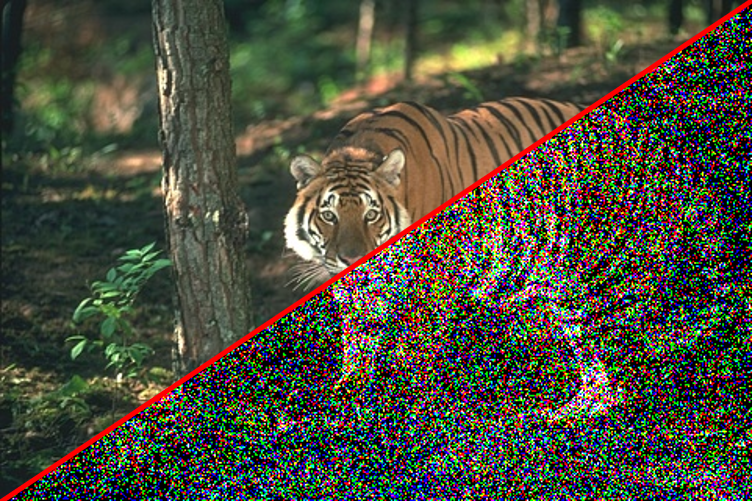}}
	\vspace{-3pt}
	\centerline{\small{(a) Ground Truth/Noisy}}
	\centerline{\small{$\lambda=1$}}
\end{minipage}
\vspace{4pt}
\begin{minipage}[h]{.49\linewidth}
	\centerline{\includegraphics[width=\linewidth]{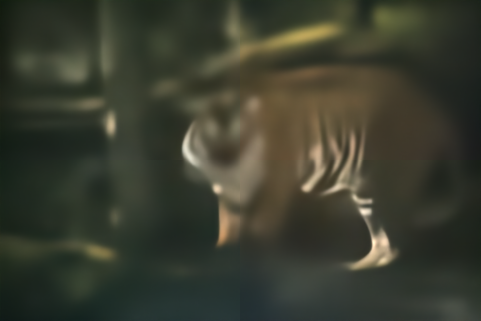}}
	\vspace{-3pt}
	\centerline{\small{(b) Anscombe (initialization)}}
	\centerline{\small{PSNR=17.89}}
\end{minipage}
\vspace{4pt}
\begin{minipage}[h]{.49\linewidth}
	\centerline{\includegraphics[width=\linewidth]{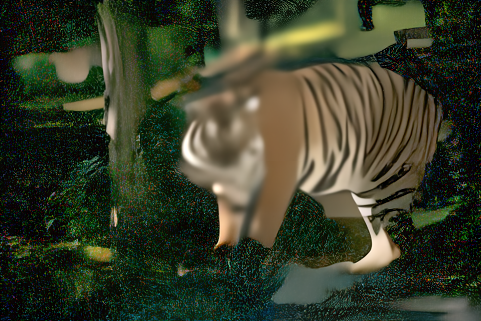}}
	\vspace{-3pt}
	\centerline{\small{(c) No Preconditioning N=100}}
	\centerline{\small{PSNR=16.90}}
\end{minipage}
\begin{minipage}[h]{.49\linewidth}
	\centerline{\includegraphics[width=\linewidth]{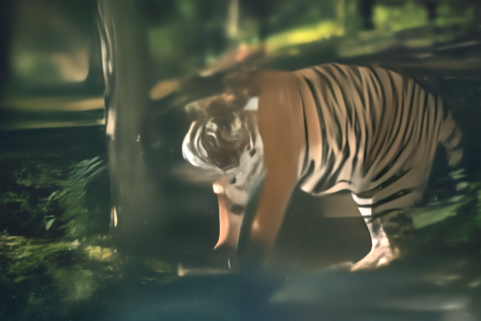}}
	\vspace{-3pt}
	\centerline{\small{(d) With Preconditioning N=100}}
	\centerline{\small{PSNR=22.30}}
\end{minipage}
\vspace{-5pt}
\caption{Poisson Denoising for very high noise level (using peak value $\lambda=1$).}
\label{fig:den-Poisson-vizu2}
\vspace{-5pt}
\end{figure}

Let us now evaluate our method for an extreme noise level, i.e. using $\lambda=1$. In this situation, the noisy image is a very inaccurate estimate of the noise variance, and it can't be used to initialize the preconditioning matrix. Instead, we compute an initial estimation using variance stabilization and Gaussian denoising: the Anscombe transform~\cite{Anscombe} is first applied to obtain an image with approximately constant noise variance; our Gaussian denoiser is then used assuming a constant standard deviation; finally the inverse Anscombe transform is applied after denoising. The preconditioning matrix $\matr{P}$ is thus computed as the square root of this initial estimate. We also disabled the update of $\matr{P}$ which degraded our results in this case. The PSNR results using this initialization method are presented in Table~\ref{tab:den-Poisson2}, and visual comparisons are shown in Fig.~\ref{fig:den-Poisson-vizu2}. Here, the best results are obtained after convergence either with or without the preconditioning. Hence the results are given for a large number of iterations $N=100$. Our approach significantly outperforms both the non-preconditioned PnP-ADMM and the Anscombe variance stabilization method used for the initialization.

\begin{table}
\centering
\renewcommand*{\arraystretch}{1.3}
\setlength\tabcolsep{3pt}
\caption{Results of PnP-ADMM for Poisson denoising either with or without preconditioning (average PSNR over each dataset).}
\vspace{-10pt}
\begin{tabular}{ccCCCC}
& & \multicolumn{2}{c}{Without Preconditioning} & \multicolumn{2}{c}{With Preconditioning} \\ \cline{3-6} 
& \multicolumn{1}{c|}{Dataset}
&  $N=6$  &  \multicolumn{1}{c|}{$N=20$} &  $N=6$  &  \multicolumn{1}{c|}{$N=20$} \\ \cline{2-6} 
\multicolumn{1}{c|}{\multirow{2}{*}{\parbox{10mm}{\rotatebox{0}{$\lambda=255$}}}}
& \multicolumn{1}{c|}{Set5}
&  36.54  &  \multicolumn{1}{c|}{36.33}  &  36.82  &  \multicolumn{1}{c|}{36.81} \\
\multicolumn{1}{c|}{}
& \multicolumn{1}{c|}{CBSD68}
&  36.35  &  \multicolumn{1}{c|}{36.25}  &  36.58  &  \multicolumn{1}{c|}{36.57} \\ \cline{2-6}
\multicolumn{1}{c|}{\multirow{2}{*}{\parbox{10mm}{\rotatebox{0}{$\lambda=\frac{255}{4}$}}}}
& \multicolumn{1}{c|}{Set5}
&  33.34  &  \multicolumn{1}{c|}{33.13}  &  33.65  &  \multicolumn{1}{c|}{33.69} \\
\multicolumn{1}{c|}{}
& \multicolumn{1}{c|}{CBSD68}
&  32.36  &  \multicolumn{1}{c|}{32.19}  &  32.72  &  \multicolumn{1}{c|}{32.74} \\ \cline{2-6}
\multicolumn{1}{c|}{\multirow{2}{*}{\parbox{10mm}{\rotatebox{0}{$\lambda=\frac{255}{8}$}}}}
& \multicolumn{1}{c|}{Set5}
&  31.73  &  \multicolumn{1}{c|}{31.41}  &  31.94  &  \multicolumn{1}{c|}{32.06} \\
\multicolumn{1}{c|}{}
& \multicolumn{1}{c|}{CBSD68}
&  30.42  &  \multicolumn{1}{c|}{30.20}  &  30.81  &  \multicolumn{1}{c|}{30.89} \\ \cline{2-6}
\end{tabular}
\vspace{-5pt}
\label{tab:den-Poisson}
\end{table}

\begin{table}
\centering
\renewcommand*{\arraystretch}{1.1}
\setlength\tabcolsep{3pt}
\caption{Denoising Results for extreme Poisson noise with $\lambda=1$ (average PSNR over each dataset).}
\vspace{-10pt}
\begin{tabular}{c|c|c|c|}
\cline{2-4}
                             & \specialcell{Anscombe\\initialization} & \specialcell{PnP-ADMM (N=100)\\no preconditioning} & \specialcell{PnP-ADMM (N=100)\\with preconditioning} \\ \hline
\multicolumn{1}{|c|}{Set5}   & 17.50 & 22.04 & 23.82 \\ \hline
\multicolumn{1}{|c|}{CBSD68} & 17.30 & 20.50 & 23.05 \\ \hline
\end{tabular}
\vspace{-5pt}
\label{tab:den-Poisson2}
\end{table}

\section{Conclusion}

In this paper, we have proposed a new preconditioning approach for Plug-and-Play optimization methods in the context of image restoration. Existing PnP schemes perform regularization of inverse problems using any conventional denoiser that assumes independent and identically distributed Gaussian noise.
On the theoretical level, we have shown that our preconditioning removes this i.i.d. assumption by introducing in the algorithm a denoiser parameterized with a covariance matrix (directly related to the chosen preconditioning matrix) instead of a single standard deviation parameter. Greater control is thus granted for better adapting the algorithm to each task by accounting for the specific error distribution of the current image estimate.

For our practical implementation, we have proposed a training procedure that generalizes a state-of-the-art CNN denoiser, enabling its parameterization with an arbitrary noise level map. While this restricts our study to diagonal covariance and preconditioning matrices, it is sufficient for many applications where the input image is degraded independently on each pixel. Such applications include image interpolation, completion, demosaicing and Poisson denoising. For each of these tasks, we have defined a suitable preconditioning scheme that significantly outperforms the non-preconditioned version both in convergence speed and image quality.
The proposed method also preserves the genericity of the PnP approach since a denoiser is used to solve several inverse problems.
Nevertheless, as we have shown with demosaicing, our method may be further specialized and improved for a given task by training the denoiser using the corresponding noise level maps.

A possible direction for future work would be to develop denoisers suitable for correlated noise (i.e. non-diagonal covariance matrix), hence extending the use of our preconditioned PnP scheme to a broader range of applications.

\ifdefined\generateAppendix
\input{appendices_content}
\fi

\section*{Acknowledgments}
This work was supported in part by the french ANR research agency in the context of the artificial intelligence project DeepCIM, and in part by the EU H2020 Research and Innovation Programme under grant agreement No 694122 (ERC advanced grant CLIM).

\bibliographystyle{IEEEtran}
\bibliography{refs}

\ifdefined\generateBiographies

\begin{IEEEbiography}
[{\includegraphics[width=1in,height=1.25in,clip,keepaspectratio]{photos/Mikael_Le_Pendu.jpg}}]%
{Mika\"el Le Pendu} received the Engineering degree from the Ecole Nationale Sup\'erieure des Mines de Nantes, Nantes, France, in 2012, and the Ph.D. degree in computer science from the University of Rennes 1, Rennes, France, in 2016. His Ph.D. studies were conducted in conjunction between the Institut National de Recherche en Informatique et en Automatique (INRIA) and Technicolor in Rennes, France, and addressed the compression of High-Dynamic Range video content. After pursuing post-doctoral research on light field image processing successively at INRIA and Trinity College Dublin, he is now a Post-Doctoral Researcher at INRIA Rennes. His research interests are image and video processing with a focus on immersive image modalities (e.g. High-Dynamic Range, Light Fields) for solving various problems including compression, super-resolution, inpainting, denoising, etc.
\end{IEEEbiography}

\begin{IEEEbiography}
[{\includegraphics[width=1in,height=1.25in,clip,keepaspectratio]{photos/photo-ChristineGuillemot.png}}]%
{Christine Guillemot}, IEEE fellow, is Director of Research at INRIA, head of a research team dealing with image and video modeling, processing, coding and communication. She holds a Ph.D. degree from ENST (Ecole Nationale Superieure des Telecommunications) Paris, and an Habilitation for Research Direction from the University of Rennes. From 1985 to Oct. 1997, she has been with FRANCE TELECOM, where she has been involved in various projects in the area of image and video coding for TV, HDTV and multimedia. From Jan. 1990 to mid 1991, she has worked at Bellcore, NJ, USA, as a visiting scientist. Her research interests are signal and image processing, and in particular 2D and 3D image and video processing for various problems (compression, super-resolution, inpainting, classification).
She has served as Associate Editor for IEEE Trans. on Image Processing (from 2000 to 2003, and from 2014-2016), for IEEE Trans. on Circuits and Systems for Video Technology (from 2004 to 2006), and for IEEE Trans. on Signal Processing (2007-2009). She has served as senior member of the editorial board of the IEEE journal on selected topics in signal processing (2013-2015) and is currently senior area editor of IEEE Trans. on Image Processing.
\end{IEEEbiography}

\fi

\end{document}

%% file: appendices_content.tex
\newcommand{\externaltext}[1]{\ifdefined\externalAppendix{#1}\fi}

\begin{appendices}
\section{Poisson data term sub-problem}
\label{app:Poisson}

Let us find the closed form solution of the minimization in Eq.~\eqref{eq:x-update-Poisson1} \externaltext{from the paper} which we rewrite here without the iteration numbers for simplicity of notation:
\begin{equation}
\label{eq_app:x-update-Poisson1}
\argmin_\vect{x} -\vect{b}^\mathsf{T}\ln(\matr{P}\vect{x}) + \mathds{1}^\mathsf{T}\vect{\matr{P}x} + \frac{\rho}{2}\norm{\vect{x}-\vect{u}}_2^2,
\end{equation}
where $\matr{P}$ is a diagonal matrix. Here, we additionally assume $\rho>0$, $\vect{b}_i\geq0$ and $\matr{P}_{i,i}>0,\,\forall i$).

Since $\matr{P}$ is diagonal, the problem is equivalently expressed for each element $i$ independently. Using the scalar notations $b=\vect{b}_i$, $p=\matr{P}_{i,i}$ and $u=\vect{u}_i$, the problem is to find the value $x$ that minimizes the function $h$ defined as:
\begin{equation}
\label{eq_app:h}
h(x)=-b\cdot\ln(x\cdot p) + x\cdot p + \frac{\rho}{2}(x-u)^2
\end{equation}
The function is defined for $x>0$ and is convex. Therefore, finding $x>0$ such that $\frac{\mathrm{d}h}{\mathrm{d}x}(x)=0$ gives the global minimum. Differentiating $h$ gives:
\begin{align}
\frac{\mathrm{d}h}{\mathrm{d}x} &= \frac{-b}{x} + p + \rho\cdot(x-u)=0,\\
&=\frac{\rho}{x}\cdot\left( x^2 + x\cdot\left(\frac{p-\rho u}{\rho}\right) - \frac{b}{\rho} \right).
\end{align}
Therefore, if we can find $x>0$ such that $x$ is a root of the second order polynomial \mbox{$x^2 + x\cdot\left(\frac{p-\rho u}{\rho}\right) - \frac{b}{\rho}$}, then $x$ minimizes the function $h$. For $b_i>0$, the solution always exists and is:
\begin{align}
    x&=\frac{1}{2}\left(\frac{\rho u-p}{\rho} + \sqrt{\left(\frac{\rho u-p}{\rho}\right)^2+\frac{4b}{\rho}}\right),\\
    \label{eq_app:xroot}
    &=\frac{1}{2\rho}\left(\rho u-p+\sqrt{(\rho u-p)^2+4\rho b}\right).
\end{align}
Note that if $b=0$ and $u\leq p/\rho$, then the highest root is \mbox{$x=0$}, which is not in the domain of $h$. However, when \mbox{$b=0$}, we can simply redefine $h$ using \mbox{$h(x)=x\cdot p + \frac{\rho}{2}(x-u)^2$} which is defined for $x=0$. So in practice, we can extend the domain of $h$ to the value $0$ (negative values are not valid pixel values and should remain excluded). With this definition, Eq.~\eqref{eq_app:xroot} gives the solution in every case.


Rewritting Eq.\eqref{eq_app:xroot} using the matrix notations directly gives the solution to Eq.~\eqref{eq_app:x-update-Poisson1}:
\begin{equation}
\label{eq_app:x-update-Poisson2}
\vect{x}_i = \frac{\rho\vect{u}_i-\matr{P}_{i,i}+\sqrt{(\rho\vect{u}_i-\matr{P}_{i,i})^2+4\rho\vect{b}_i}}{2\rho}.
\end{equation}


\end{appendices}